\pdfoutput=1
\documentclass[PAPER, atlasdraft=false, cernpreprint, UKenglish, texmf, orcidlogo]{atlasdoc}
\usepackage{atlaspackage}
\usepackage{atlasbiblatex}

\usepackage{atlasphysics}
\ifthenelse{\boolean{AtlasDraft}}{\usepackage[output=true,
shift=true]{atlastodo}}{}

\graphicspath{{logos/}{figures/}}

\usepackage{ANA-HIGP-2024-16-PAPER-defs}

\AtlasTitle{Measurement of the Higgs boson decay to a low-mass dilepton
system and a photon in $pp$ collisions at $\sqrt{s} =$ 13 and 13.6 TeV
with the ATLAS detector}

\AtlasAbstract{%
A measurement is presented for the Higgs boson decaying into a photon
and a pair of electrons or muons with an invariant mass
$m_{\ell\ell} <$ 30 GeV,
using $pp$ collision data at $\sqrt{s} =$ 13.6 TeV recorded by the
ATLAS detector at the Large Hadron Collider during the years 2022--2024,
corresponding to an integrated luminosity of 164 fb$^{-1}$.
The best-fit value of the signal strength parameter, defined as the
ratio of the observed signal yield to the one expected in the
Standard Model, is measured to be $\mu = 0.64^{+0.44}_{-0.39}$
compared with the expected value of $\mu = 1.00^{+0.47}_{-0.41}$.
The corresponding observed (expected) signal significance is 1.7
(2.5) standard deviations under the background-only hypothesis.
The result is combined with a similar search performed using
139 fb$^{-1}$ of $pp$ collision data at $\sqrt{s} =$
13 TeV collected during the years 2015--2018.
This combination gives an observed (expected) signal strength of $\mu =
1.03^{+0.35}_{-0.32} (\mu = 1.00^{+0.34}_{-0.32})$, corresponding
to an observed (expected) significance of 3.4 (3.3) standard
deviations.
}

\AtlasRefCode{HIGP-2024-16}

\PreprintIdNumber{CERN-EP-2026-223}

\AtlasJournal{JHEP}
\AtlasCoverEgroupAnalysisTeam{atlas-higp-2024-16-analysis-team@cern.ch}

\hypersetup{pdftitle={ATLAS document},pdfauthor={The ATLAS Collaboration}}

\begin{document}

\maketitle

\tableofcontents
\newpage

%
%

%

\section{Introduction}
\label{sec:intro}

In 2012, the ATLAS and CMS Collaborations at the Large Hadron
Collider (LHC)~\cite{Evans:2008zzb} discovered a new particle with a mass
around \qty{125}{\GeV}, consistent with the Higgs boson predicted by
the Standard Model (SM)~\cite{HIGG-2012-27, CMS-HIG-12-028}.
Measurements of the particle's properties, such as spin, parity,
and its couplings to other SM particles, show no significant
deviations from the SM predictions~\cite{HIGG-2021-23,
CMS-HIG-22-001, HIGG-2013-17, CMS-HIG-14-018}.

Within the SM, the Higgs boson decay to a low-mass dilepton system
and a photon ($H\to\ell\ell\gamma$, where $\ell = e, \mu$)
proceeds mainly through loop-induced processes.
The branching ratio for $H\to\ell\ell\gamma$ is approximately
5\% of that for $H\to\gamma\gamma$~\cite{Firan:2007tp},
making this decay particularly challenging to observe experimentally.
The ratio of the $H\to\gamma^{*}\gamma\to\ell\ell\gamma$ and
$H\to\gamma\gamma$ decay rates can also provide a complementary
test of the loop-induced Higgs–photon interaction that is largely
independent of the Higgs boson production cross-section.
The non-zero virtuality of the $\gamma^{*}$ also provides sensitivity to possible
new physics effects not accessible in the $H\to\gamma\gamma$ decay~\cite{Kachanovich:2020xyg}.
These three-body decays can also be used to probe the CP properties of the Higgs
boson by studying the forward–backward asymmetry of its decay
products~\cite{Korchin:2014kha, Chen:2014ona}.

\begin{figure}[!htp]
\centering
\subfloat[]{\includegraphics[width=0.28\textwidth]{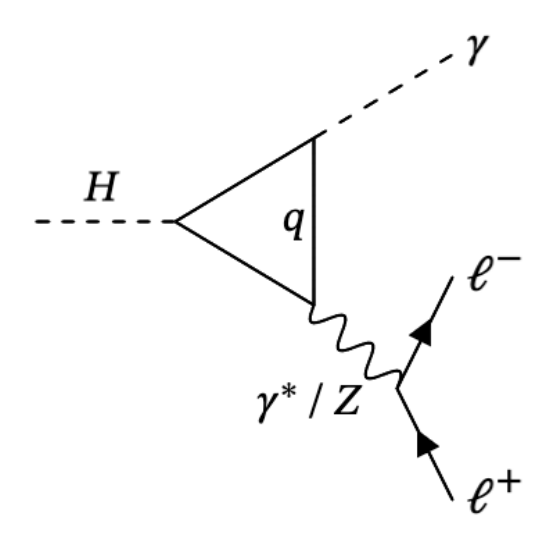}}
\subfloat[]{\includegraphics[width=0.28\textwidth]{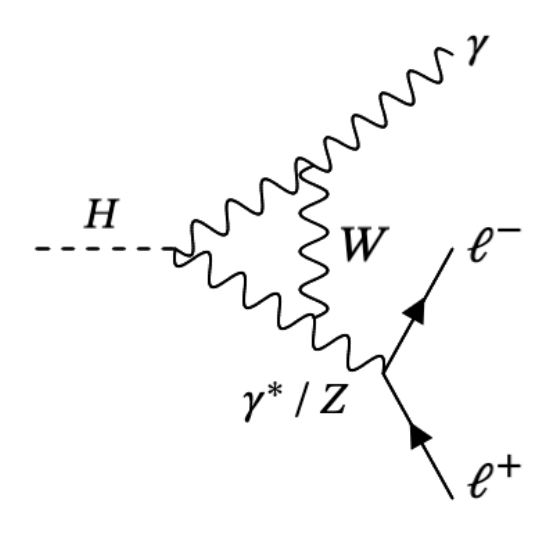}}
\subfloat[]{\includegraphics[width=0.28\textwidth]{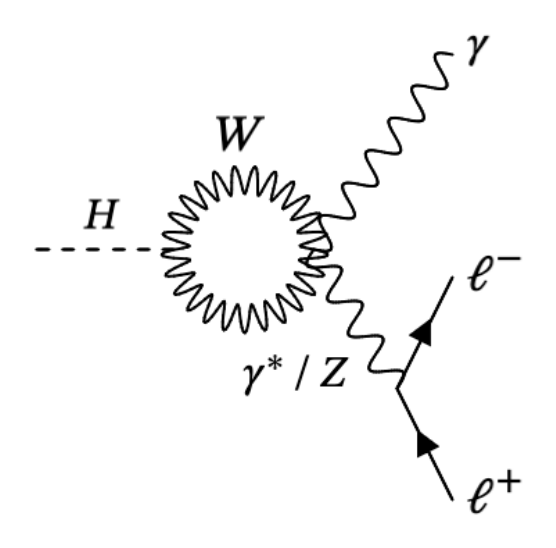}}
\caption{Representative Feynman diagrams contributing to the loop-induced
$H\to V^*\gamma \to \ell\ell\gamma$ decay,
where $V^{*} = \gamma^{*}$ or $Z$.
Diagram (a) shows the fermion loop contribution, while (b) and
(c) show electroweak gauge boson loop contributions.}
\label{fig:feynman-diagrams}
\end{figure}

The region considered in this analysis, $m_{\ell\ell} < \qty{30}{\GeV}$,
is dominated by the loop-induced $H\to \gamma^{*}\gamma\to \ell\ell\gamma$ decay,
illustrated in Figure~\ref{fig:feynman-diagrams}~\cite{Kachanovich:2020xyg, Chen:2012ju, Passarino:2013nka}.
Contributions involving a $Z$ boson ($H\to Z\gamma\to \ell\ell\gamma$), including interference effects
, are below 1\% and therefore
neglected~\cite{Kachanovich:2021pvx}.

Both the ATLAS and CMS Collaborations have performed searches for the
$H \to \ell\ell\gamma$ decay at low $m_{\ell\ell}$ using
Run-2 proton--proton ($pp$) collision
data at $\sqrt{s} = \qty{13}{\TeV}$.
The ATLAS Collaboration reported evidence for the $H\to
\ell\ell\gamma$ process
using \qty{139}{\ifb} of data collected in the years 2015--2018 (Run 2), with
a significance of 3.2$\sigma$ over the background-only hypothesis
compared with an expected significance of 2.1$\sigma$ for the SM
prediction~\cite{HIGG-2018-43}.
The signal strength parameter, defined as the ratio of the observed
signal yield to the SM prediction, was measured to be $\mu = 1.5 \pm
0.5$~\cite{HIGG-2018-43}.
The CMS Collaboration set a 95\% CL upper limit on the
production cross-section times branching ratio for the
low-$m_{\mu\mu}$ $H\to \mu\mu\gamma$ channel of 4.0 times the
SM prediction, using \qty{35.9}{\ifb} of data collected in
2016~\cite{CMS-HIG-17-007}.

Searches for $H\to Z\gamma$, which target the complementary
region with $m_{\ell\ell} > \qty{30}{\GeV}$, have been performed by
both the ATLAS
and CMS Collaborations at $\sqrt{s} = \qty{13}{\TeV}$
and $\qty{13.6}{\TeV}$~\cite{HIGG-2018-43, HIGG-2018-42,
CMS-HIG-19-014, HIGG-2022-22, HIGP-2024-19}.
The most recent ATLAS search reports an observed (expected)
significance of 2.5 (1.9)$\sigma$, obtained from the combination of
\qty{165}{\ifb} data at $\sqrt{s} = \qty{13.6}{\TeV}$ collected in 2022--2024 (Run 3)
with the Run-2 data of \qty{139}{\ifb} at $\sqrt{s} = \qty{13}{\TeV}$.
The ATLAS and CMS Collaborations additionally combined their Run-2
results and provided the first evidence for the $H \to Z\gamma$ decay
with a statistical significance of
3.4 $\sigma$~\cite{HIGG-2022-22}.

This paper presents a measurement of the $H\to \ell\ell\gamma$
decay in the region $m_{\ell\ell} <
\qty{30}{\GeV}$.
The measurement uses $pp$ collision data at $\sqrt{s} = \qty{13.6}{\TeV}$
collected by the ATLAS detector in 2022--2024, corresponding to a total integrated luminosity of \qty{164}{\ifb}.
Events are divided into nine mutually exclusive categories based on event
topology and lepton flavour.
In all categories, the dominant background comes from
irreducible non-resonant $\ell\ell\gamma$ production. The signal
yield is extracted through a simultaneous fit of parametric
functions to the reconstructed $\ell\ell\gamma$ invariant
mass ($m_{\ell\ell\gamma}$) distributions across all categories.

Due to the low invariant mass of the two leptons and high
transverse momentum of the dilepton system, in the $ee\gamma$
categories the electromagnetic showers associated with the electron
candidates can overlap (merge) in the calorimeter.
To address this, dedicated electron trigger and identification
algorithms were developed to better reconstruct and identify
$ee\gamma$ events with merged dielectron candidates. The
identification algorithm for merged dielectrons presented in this
paper is based on a boosted decision tree (BDT) and improves the final expected significance by 5\% relative to
the cut-based identification used in Ref.~\cite{HIGG-2018-43},
when evaluated at the same cross-section and luminosity.
The analysis furthermore benefits from the higher Higgs-boson production cross-section
at the increased centre-of-mass energy and from the larger data set.


%

%

\section{ATLAS detector}
\label{sec:detector}

The ATLAS experiment~\cite{PERF-2007-01,GENR-2019-02} at the LHC is a multipurpose particle detector with a forward--backward symmetric cylindrical geometry and near $4\pi$ coverage in solid angle\footnote{ATLAS uses a right-handed coordinate system
with its origin at the nominal interaction point (IP) in the centre of the detector and the $z$-axis along the beam pipe.
The $x$-axis points from the IP to the centre of the LHC ring, and the $y$-axis points upwards.
Polar coordinates ($r,\phi$) are used in the transverse plane, $\phi$ being the azimuthal angle around the $z$-axis.
The pseudorapidity is defined in terms of the polar angle $\theta$ as $\eta = -\ln\tan(\theta/2)$ and is equal to the rapidity $y = \frac{1}{2}\ln\left(\frac{E+p_{z}}{E-p_{z}}\right)$
in the relativistic limit.
Angular distance is measured in units of $\Delta R \equiv \sqrt{(\Delta y)^{2}+(\Delta \phi)^{2}}$.}.
It consists of an inner tracking detector (ID) surrounded by a thin superconducting solenoid providing a \qty{2}{\tesla} axial magnetic field, electromagnetic (EM) and hadronic calorimeters, and a muon spectrometer (MS).
The inner tracking detector covers the pseudorapidity range $|\eta| <$ 2.5 and consists of silicon pixels, silicon microstrips, and transition radiation tracking detectors.
Lead/liquid-argon (LAr) sampling calorimeters provide electromagnetic energy measurements with high granularity within the region $|\eta| <$ 3.2.
A steel/scintillator-tile hadronic calorimeter covers the central pseudorapidity range ($|\eta| <$ 1.7).
The endcap and forward regions are instrumented with LAr calorimeters for EM and hadronic energy measurements up to $|\eta| =$ 4.9.
The muon spectrometer surrounds the calorimeters and is based on three large superconducting air-core toroidal magnets with eight coils each.
The muon spectrometer includes a system of precision tracking chambers up to $|\eta| =$ 2.7 and fast detectors for triggering up to $|\eta| =$ 2.4.
The luminosity is primarily measured by the LUCID-2 detector, which is located close to the beam pipe.

A two-level trigger system was used to select events~\cite{TRIG-2016-01,TRIG-2022-01}.
The first-level trigger is implemented in hardware and uses a subset of the detector information to accept events at a rate close to \qty{100}{\kilo\hertz}.
This is followed by a software-based trigger that reduces the accepted rate of complete events to \qty{3}{\kilo\hertz} on average, depending on the data-taking conditions.
A software suite~\cite{SOFT-2022-02} is used in data simulation, in the reconstruction and analysis of real and simulated data, in detector operations, and in the trigger and data acquisition systems of the experiment.


%

%

\section{Data and simulation samples}
\label{sec:samples}

The data was collected at $\sqrt{s} = \qty{13.6}{\TeV}$ in 2022--2024,
using a combination of single-lepton, dilepton,
single-photon, diphoton, and lepton$+$photon
triggers~\cite{TRIG-2018-05, TRIG-2012-03}, with a variety of $p_\mathrm{T}$ thresholds
chosen to maximise signal efficiency.
The lowest-threshold single-electron and single-muon triggers required
$p_\mathrm{T} > \qty{26}{\GeV}$ and $p_\mathrm{T} > \qty{24}{\GeV}$, respectively.
The dielectron trigger required two electrons with $p_\mathrm{T} >
\qty{17}{\GeV}$ each.
The dimuon trigger used a symmetric 14--\qty{14}{\GeV}
configuration, and in the years 2023--2024 an asymmetric
22--\qty{8}{\GeV} configuration was also available.
The lepton+photon triggers required a photon with
$p_\mathrm{T} > \qty{25}{\GeV}$ in association with either an electron or a
muon with $p_\mathrm{T} > \qty{24}{\GeV}$.
The single-photon trigger required $p_\mathrm{T} >
\qty{140}{\GeV}$, while the diphoton trigger required two photons
with transverse energies of at least \qty{35}{\GeV}
(\qty{25}{\GeV}) for the leading (subleading) photon\footnote{
The diphoton trigger collects $ee\gamma$ events since resolved electrons and merged
dielectrons can satisfy the photon trigger requirements.
}.
A dedicated trigger for merged dielectron objects was also used~\cite{TRIG-2018-05}.
To improve the efficiency at high instantaneous luminosity, these
low-threshold triggers were supplemented by higher $p_\mathrm{T}$ triggers
with looser identification or isolation requirements.
The overall trigger efficiency for signal events is approximately
92\% for events passing all offline analysis selections.
After data quality requirements, the total integrated luminosity
amounts to \qty{164}{\ifb}.
The average number of inelastic $pp$ interactions per bunch crossing
(pile-up) increased from 42 in 2022 to 58 in 2024, with the peak
instantaneous luminosity reaching
\SI{2.3e34}{\per\centi\meter\squared\per\second}~\cite{DAPR-2018-01}.

Simulated Monte Carlo (MC) samples are used to optimise the event selection,
model the signal processes, and build templates for the background modelling studies.
The signal samples were produced using the full ATLAS detector simulation~\cite{SOFT-2010-01},
based on \GEANT~\cite{Agostinelli:2002hh},
while the ATLAS fast simulation (AF3)~\cite{SIMU-2018-04}
was employed for the dominant background to enable large samples to be produced.
The effect of pile-up was modelled by overlaying~\cite{SIMU-2020-01}
the simulated hard-scattering event with inelastic $pp$ events generated from a mix of
\EPOS[2.0.1.4]~\cite{Werner:2005jf} with the \EPOS LHC tune~\cite{Pierog:2013ria}
and \PYTHIA[8.308]~\cite{Bierlich:2022pfr} with the A3 tune~\cite{ATL-PHYS-PUB-2016-017}.
The pile-up modelling in simulation was corrected by reweighting the simulated
events so that the distribution of the average number of interactions per bunch
crossing matched that observed in data.
The simulation was further corrected with scale factors accounting for residual
differences from data in the reconstruction, identification and isolation
efficiencies of the reconstructed objects, and in their energy and momentum
scales and resolutions.

Higgs boson production via gluon--gluon fusion
(ggF)~\cite{deFlorian:2016spz,
Anastasiou:2015ema, Anastasiou:2016cez, Dulat:2018rbf,
Harlander:2009mq, Harlander:2009bw, Harlander:2009my, Pak:2009dg,
Actis:2008ug, Actis:2008ts, Anastasiou:2008tj, Aglietti:2004nj,
Bonetti:2018ukf},
vector-boson fusion
(VBF)~\cite{deFlorian:2016spz,
Ciccolini:2007jr, Ciccolini:2007ec, Bolzoni:2010xr},
as well as quark-initiated associated production with a vector boson
($VH$, where $V = W,
Z$)~\cite{deFlorian:2016spz, Ciccolini:2003jy,
Brein:2003wg, Brein:2011vx, Altenkamp:2012sx, Denner:2014cla,
Brein:2012ne, Harlander:2014wda, Harlander:2018yio}
or with two top quarks
($t\bar{t}H$)~\cite{deFlorian:2016spz,
Beenakker:2002nc, Dawson:2003zu, Zhang:2014gcy, Frixione:2015zaa}
was modelled using the \POWHEGBOX[v2] MC event
generator~\cite{Nason:2009ai, Alioli:2010xd, Nason:2004rx,
Frixione:2007vw, Hartanto:2015uka}.
The $H\to \gamma^{*}\gamma \to \ell\ell\gamma$ decay, parton showering,
hadronisation, and the underlying event were modelled with \PYTHIA[8]~\cite{Bierlich:2022pfr}
for all samples.
The Higgs boson mass was set to $m_{H} =
\qty{125}{\GeV}$ and its total decay width was set to $\Gamma_{H} =
\qty{4.07}{\MeV}$~\cite{deFlorian:2016spz}.
The samples are normalised to the SM production cross-sections,
evaluated at a Higgs boson mass of $m_{H} =
\qty{125.09}{\GeV}$~\cite{Karlberg:2024zxx}.
The branching ratios are taken as
$\mathcal{B}(H\to ee\gamma) = 7.20\times 10^{-5}$ and
$\mathcal{B}(H\to\mu\mu\gamma) = 3.42\times 10^{-5}$ for
$m_{\ell\ell} < \qty{30}{\GeV}$~\cite{Firan:2007tp}.
Currently, no theoretical uncertainty calculation exists for the
low-$m_{\ell\ell}$ $H\to\ell\ell\gamma$ branching ratio.
Therefore, the theoretical uncertainty of 6.9\% on the $H\to
Z\gamma$ branching ratio is used~\cite{HIGP-2024-19,
deFlorian:2016spz}.
The ggF sample was generated at next-to-next-to-leading order (NNLO)
accuracy in quantum chromodynamics (QCD),
and normalised to the next-to-next-to-next-to-leading order
(N$^{3}$LO) in QCD plus electroweak
corrections (EW) at next-to-leading order (NLO) cross-section prediction.
The VBF, $VH$, and $t\bar{t}H$ samples were generated at NLO accuracy,
and subsequently
normalised to approximate-NNLO QCD + NLO EW cross-section for VBF,
NNLO QCD + NLO EW for $VH$,
and NLO QCD + NLO EW cross-section for $t\bar{t}H$.
The loop-induced ($ggZH$) contribution was generated separately at leading-order (LO)
QCD and then normalised to NLO + next-to-leading logarithmic (NLL) QCD prediction.
No dedicated sample was generated for Higgs boson production in
association with two
$b$-quarks ($b\bar{b}H$), as it contributes only about 1\% of the
total expected Higgs boson production.
The effect is taken into account by scaling the ggF sample according
to the total ggF+$b\bar{b}H$ cross-section, and a 50\% uncertainty is
assigned to the $b\bar{b}H$ yield.
The uncertainty associated with the parton shower modelling of the
$H\to \gamma^{*}\gamma$ signal is estimated using $H\to\gamma\gamma$
samples showered with
\textsc{Herwig7}~\cite{Bellm:2019zci} for the ggF and VBF production modes.

The dominant background comes from non-resonant $\ell\ell\gamma$ production.
Events were simulated with the \SHERPA[2.2.14]~\cite{Bothmann:2019yzt}
generator based on LO matrix elements for
$\ell\ell\gamma$ production with up to three additional partons
and using the NNPDF3.0NNLO PDF set~\cite{Ball:2014uwa}.
The \SHERPA simulation includes parton shower, fragmentation, and
underlying-event modelling.

The smaller resonant background component from $H\to\gamma\gamma$ events is also
considered.
This background arises when one photon converts and passes the $H\to ee\gamma$ selection criteria.
A configuration identical to that used to generate
signal samples was used to generate samples with $H
\to \gamma \gamma$ decays.


%

%

\section{Object reconstruction and event selection}
\label{sec:selection}

Events are required to have at least one primary vertex with at least two associated tracks with
$p_\mathrm{T} > \qty{0.5}{\GeV}$.
The vertex with the largest $\sum
p_\mathrm{T}^{2}$ of the associated tracks is considered to be the primary
vertex~\cite{IDTR-2021-01}.

\subsection{Object reconstruction}
Photon candidates are reconstructed by clustering EM calorimeter cells
with significant energy deposits into ``topological clusters'',
based on the significance of energy deposits~\cite{EGAM-2021-01}.
The photon candidates are required to satisfy the \textit{Loose} identification criteria~\cite{EGAM-2018-01, EGAM-2021-01},
$p_\mathrm{T} >
\qty{20}{\GeV}$ and $|\eta| <$ 2.37, excluding the transition region between
the barrel and endcap EM calorimeters (1.37 $< |\eta| <$ 1.52).
They are labelled as \textit{converted} if
matched either to two tracks from one conversion vertex or to a
single track with the signature of an electron track but no hit in
the innermost pixel layer of the inner detector; otherwise, they are
labelled as \textit{unconverted}. About 30\% of
photon candidates are classified as converted photons.

Electrons are reconstructed by matching ID tracks to topological
calorimeter energy clusters formed using the same dynamical cell-clustering
algorithm as in the photon reconstruction~\cite{EGAM-2021-01}.
Because of the event kinematics of the $H\to \gamma^{*}\gamma
\to \ell\ell\gamma$ signal,
it is common for the energy deposits of the two electrons in the EM
calorimeter to be reconstructed as a single cluster.
Therefore, two types of electron candidates are considered:
resolved electrons~\cite{EGAM-2021-01} with a single ID track matched to a topological cluster,
and merged dielectron where two ID tracks are matched to one topological cluster.
Resolved electron candidates are preselected by satisfying
$p_\mathrm{T} > \qty{4.5}{\GeV}$, $|\eta| <$ 2.47, and the
\textit{VeryLoose} selection criteria,
based on a likelihood discriminant derived from calorimeter shower
shapes and track parameters~\cite{EGAM-2021-01}.
Merged dielectron candidates are required to have $p_\mathrm{T} >
\qty{4.5}{\GeV}$ and $|\eta| <$ 2.37.
The tracks considered must have opposite charge,
$p_\mathrm{T} > \qty{0.5}{\GeV}$, $|\eta| <$ 2.5, and at least seven hits in
the silicon detector as well as a hit in the innermost pixel layer to
suppress converted-photon background.
If more than two opposite-charge tracks pass these requirements, the
pair with the largest di-track $\pt$ is selected.
The mass and direction of the dielectron object is taken from the di-track vector sum.
For both types, electron candidates in the transition region between the
barrel and endcap EM calorimeters (1.37 $< |\eta| <$ 1.52) are excluded.
In addition, each ID track must
be matched to the primary vertex via a requirement on the longitudinal impact
parameter $\Delta z_{0}$: $|\Delta z_{0} \cdot
\sin\theta| < \qty{0.5}{\mm}$,
where $\theta$ is the polar angle of the track.
The significance of the transverse impact parameter $d_{0}$
calculated relative to the beam-line position is required to
be $|d_{0}|/\sigma_{d_{0}} <$ 5,
where $\sigma_{d_{0}}$ is the uncertainty in $d_{0}$.

The energy of the merged dielectron object is calibrated as that of a converted
photon, since its behaviour in the calorimeter closely resembles that of a photon converting
in the detector material near the interaction vertex~\cite{HIGG-2018-43}.
A BDT is trained using
LightGBM~\cite{LightGBM_Doc} to separate merged dielectrons from the signal against photons
from $H\to\gamma\gamma$ decay and jets that are reconstructed as merged dielectrons.
The BDT uses both the standard electron-identification
shower-shape variables~\cite{EGAM-2021-01} and track-to-calorimeter
pointing variables. The track-to-calorimeter pointing variables used are:
\begin{itemize}
\item ratio between the cluster energy and the magnitude of
the momentum of the two-track system,
$E/|\vec{p}^{\,\text{trk}}_{1} + \vec{p}^{\,\text{trk}}_{2}|$
\item difference between $\eta$ of the
leading track extrapolated to the first layer of the EM
calorimeter and $\eta$ of the calorimetric cluster in the first
layer of the EM calorimeter
\item difference between $\eta$ of the
calorimetric cluster in the second layer of the EM calorimeter
and $\eta$ of the two-track system extrapolated to the calorimeter
\item difference between $\phi$ of the
calorimetric cluster in the second layer of the EM calorimeter
and $\phi$ of the two-track system extrapolated to the calorimeter
\item difference between the longitudinal
impact parameters of the two tracks
\end{itemize}
The BDT working point is chosen to match the background efficiency of
the cut-based identification algorithm used in the Run-2
analysis~\cite{HIGG-2018-43}, which ranges from 10-20\% depending on $p_\mathrm{T}$.
At the same background efficiency, the signal efficiency increases from approximately 62\%
for the cut-based method to 90\% for the BDT-based identification.
Using converted photons from $Z\to\ell\ell\gamma$ decays as
a proxy for merged dielectrons,
the scale factor for the BDT ID efficiency in simulation, together with its uncertainty,
is determined using a data-driven method.

Muon candidates are reconstructed by matching tracks in the muon
spectrometer and inner detector.
The preselected muon candidates are required to satisfy the
\textit{Medium} identification criteria~\cite{MUON-2018-03}, and have $p_\mathrm{T} > \qty{5}{\GeV}$ and $|\eta| <$ 2.7.
Muon candidates with an associated ID track must be matched to the
primary vertex through the requirements $|\Delta z_{0} \cdot
\sin\theta| < \qty{0.5}{\mm}$ and $|d_{0}|/\sigma_{d_{0}} <$ 3.

Jets are reconstructed from particle flow objects~\cite{PERF-2014-07}
clustered via the anti-$k_{t}$ algorithm ($R =$
0.4)~\cite{Cacciari:2008gp, Fastjet}.
Jet candidates are required to have $p_\mathrm{T} > \qty{20}{\GeV}$ and $|y| <$ 4.4.
Jets originating from pile-up collisions in the range $|\eta| <$ 2.5
and $p_\mathrm{T} < \qty{60}{\GeV}$ are suppressed using a jet vertex
tagging algorithm~\cite{PERF-2014-03}.
In the forward region ($|\eta| >$ 2.5), pile-up jets with
$p_\mathrm{T} < \qty{60}{\GeV}$ are suppressed using the forward jet
vertex tagging algorithm~\cite{ATL-PHYS-PUB-2019-026}.

\subsection{Event selection}
The $Z/\gamma^{*}$ candidates are selected from the preselected objects.
If there are lepton pairs of different flavours in one event, muon
pairs are given the highest priority.
If no opposite-charge muon pair satisfying the requirements above is
found, the electron-pair candidates are considered.
The resolved electron pair or merged $ee$ object with the highest
vector-sum of the $\pt$ of the associated ID tracks is selected.
The highest $\pt$ preselected photon not overlapping with the selected
$Z/\gamma^{*}$ candidate within $\Delta R <$ 0.02 is kept, and all other photon candidates are discarded.
Each candidate therefore contains one selected photon and either two opposite-charge, same-flavour leptons
(muons or resolved electrons) or one merged dielectron candidate.
Events are rejected if the photon is within a $\Delta R =$ 0.4 cone
around either of the preselected leptons.
Jets within $\Delta R =$ 0.4 of the photon are discarded.
Jets within $\Delta R =$ 0.2 of the electrons are also discarded.
If the electron-jet angular separation is 0.2 $< \Delta
R <$ 0.4, the electron, and therefore the event, is discarded.
The final identification, isolation, and kinematic requirements are applied to the objects
forming the selected candidate.
The selected photon is required to satisfy \textit{Tight}
identification and \textit{Loose} isolation
requirements~\cite{EGAM-2018-01, EGAM-2021-01}.
The leading muon (with highest $p_\mathrm{T}$) is further required to
satisfy $p_\mathrm{T} > \qty{15}{\GeV}$.
The leading muon is also required to be isolated from additional
activity in the tracking detector and in the calorimeters, using a
\textit{Tight} isolation working point with an additional
correction for close-by objects~\cite{MUON-2018-03}.
Similarly, for $ee\gamma$ events, the leading resolved electron is required to
satisfy $p_\mathrm{T} > \qty{18}{\GeV}$ and the merged dielectron object
is required to have $p_\mathrm{T} > \qty{20}{\GeV}$.
Both resolved electrons are required to satisfy the \textit{Medium}
identification criteria, while the leading one is additionally required
to satisfy the \textit{Loose} isolation requirement, with the additional
isolation correction for close-by objects; the merged dielectrons are
required to satisfy \textit{Tight} isolation requirements.
In both cases the standard electron calorimeter- and track-based
isolation working points~\cite{EGAM-2018-01} are used.

To suppress events involving $H\to Z\gamma$ decays, the dilepton
invariant mass is required to satisfy $m_{\ell\ell} < \qty{30}{\GeV}$.
Events with a $J/\psi$ are removed by excluding events where the muon
pair lies within $\qty{2.9}{\GeV} < m_{\mu\mu} < \qty{3.3}{\GeV}$
or where the electron pair lies within
$\qty{2.5}{\GeV} < m_{ee} < \qty{3.5}{\GeV}$.
Additionally, events with an $\Upsilon$, defined as muon (electron) pairs in
the mass range of $\qty{9.1}{\GeV} <
m_{\mu\mu} < \qty{10.6}{\GeV}$ ($\qty{8.0}{\GeV} < m_{ee} < \qty{11.0}{\GeV}$), are discarded.
The invariant mass of the three-body $\ell\ell\gamma$ system is required to fall within
$\qty{110}{\GeV} < m_{\ell\ell\gamma} < \qty{160}{\GeV}$.
Events are further required to satisfy the relative transverse
momentum requirements
$p_\mathrm{T}^{\gamma}/m_{\ell\ell\gamma} >$ 0.3 and
$p_\mathrm{T}^{\ell\ell}/m_{\ell\ell\gamma} >$ 0.3.

The overall signal selection efficiency, calculated relative to the
generated $H\to\gamma^{*}\gamma\to\ell\ell\gamma$ events with $\ell = e,\mu$ and
$2m_{\ell} < m_{\ell\ell} <$ \qty{90}{\GeV}, is 12.5\%
after passing all event selections.


%

%

\section{Event categorisation}
\label{sec:categorization}

To maximise the sensitivity to the $H\rightarrow \gamma^{*}\gamma
\rightarrow \ell\ell\gamma$ signal, the selected events are
classified into mutually exclusive categories, following the strategy used in
the previous publication~\cite{HIGG-2018-43}.
The categorisation is based on the lepton type and the event topology.
Three decay channels are defined according to lepton type: $\mu\mu$,
resolved $ee$ and merged $ee$, each of which is further divided into
three event topology categories.

\begin{itemize}
\item \textbf{VBF category}: events must contain at least two
jets with $p_\mathrm{T} > \qty{25}{\GeV}$.
Jets in the forward region ($|\eta_{j}| >$ 2.5) are additionally
required to satisfy $p_\mathrm{T} > \qty{30}{\GeV}$ to suppress events
with pile-up jets.
The invariant mass of the two leading jets $m_{jj}$
must be greater than \qty{500}{\GeV}.
The pseudorapidity separation between the leading jets
$|\Delta\eta_{jj}|$ is required to be greater than 2.7.
The Zeppenfeld variable, $\eta_{_{\text{Zeppenfeld}}} =
|\eta_{\ell\ell\gamma} - 0.5(\eta_{j1} + \eta_{j2})|$~\cite{PhysRevD.54.6680},
measures the centrality of the $\ell\ell\gamma$ system with
respect to the two leading jets and is required to be less than 2.0.
This exploits the vector-boson fusion topology where the $\ell\ell\gamma$ system is typically produced centrally between the two forward jets,
enhancing the signal-to-background ratio.
The minimum $\Delta R$ between either of the two
leading jets and any selected lepton or photon object must be greater
than 1.5. Finally, the $|\Delta \phi|$ between the $\ell\ell\gamma$
system and the dijet system must be greater than 2.8.
\item \textbf{High $p_\mathrm{Tt}$ category}: events failing to meet the
VBF selection but satisfying $p_\mathrm{Tt} > \qty{100}{\GeV}$ are
placed into the high $p_\mathrm{Tt}$ category. The $p_\mathrm{Tt}$ variable is
defined as the transverse momentum
projection onto the $\ell\ell\gamma$ thrust axis,
$p_\mathrm{Tt} = |\vec{p}_\mathrm{T}^{\ell\ell\gamma} \times \hat{t}|$, where
$\hat{t} = (\vec{p}_\mathrm{T}^{\ell\ell} - \vec{p}_\mathrm{T}^{\gamma}) /
|\vec{p}_\mathrm{T}^{\ell\ell} - \vec{p}_\mathrm{T}^{\gamma}|$. High $p_\mathrm{Tt}$
events exhibit harder kinematics and a better
signal-to-background separation~\cite{HIGG-2018-43}.
\item \textbf{Low $p_\mathrm{Tt}$ category}: all remaining events (failing
the VBF and high $p_\mathrm{Tt}$ selections).
\end{itemize}

The selections for the categorisation strategy are summarised in
Table~\ref{tab:category_selections}.
The analysis therefore has nine categories: three event-topology categories for each of the three decay channels.

\begin{table}[!htp]
\centering
\caption{Selections used to define the VBF, high
$p_\mathrm{Tt}$ and low $p_\mathrm{Tt}$ categories in each final state channel.}
\begin{tabular}{@{}p{3cm} p{11cm}@{}}
\toprule
Category             &   Selections                                      \\
\midrule
\multirow{7}{*}{VBF} &   Jet multiplicity $\geq$ 2, with $p_\mathrm{T}^{j} >
\qty{25}{\GeV}$     \\
&   $p_\mathrm{T}^{j} > \qty{30}{\GeV}$ if $|\eta_{j}| >$ 2.5              \\
&   $m_{jj} > \qty{500}{\GeV}$                              \\
&   $|\Delta \eta_{jj}| >$ 2.7                       \\
&   $|\eta_{\ell\ell\gamma} - 0.5(\eta_{j1} + \eta_{j2})| <$ 2.0    \\
&   $\Delta R(j_{i}, \ell) >$ 1.5,  $\Delta R(j_{i}, \gamma) >$
1.5, $i=$1,2       \\
&   $|\Delta \phi(\ell\ell\gamma, jj)| >$ 2.8       \\
\midrule
High $p_\mathrm{Tt}$        &   Failing the VBF selection  and $p_\mathrm{Tt}
> \qty{100}{\GeV}$         \\
\midrule
Low $p_\mathrm{Tt}$         &   Failing the VBF selection  and $p_\mathrm{Tt}
\le \qty{100}{\GeV}$         \\
\bottomrule
\end{tabular}
\label{tab:category_selections}
\end{table}


%

%

\section{Signal and background modelling}
\label{sec:modelling}

To extract the observed signal yield, parametric functions are used
to model the $\ell\ell\gamma$ invariant mass distributions for both
signal and background in each of the nine categories.
A combined model is constructed from these functions and
simultaneously fitted to the selected data in the
$m_{\ell\ell\gamma}$ range of 110--\qty{160}{\GeV}.

\subsection{Signal modelling}
The $H\to \gamma^{*}\gamma$ signal shape is described by a
double-sided Crystal Ball (DSCB) function,
consisting of a  Gaussian core with power-law tails on both sides to
account for detector resolution and non-Gaussian effects~\cite{HIGG-2014-04}.
The DSCB shape is determined separately in each of the nine categories after
applying the event selection described in Section~\ref{sec:selection}.
In each category, the DSCB parameters (mean $\mu_{\text{CB}}$, width
$\sigma_{\text{CB}}$, and four parameters describing the tails) are
obtained from fits to the signal samples, combining all Higgs production modes.
Since the signal samples are generated with a Higgs boson mass of
$m_{H} = \qty{125}{\GeV}$, a shift of \qty{0.09}{\GeV} is applied to the position
of the signal peak to account for the difference from the reference
Higgs boson mass used in this analysis, $m_H = \qty{125.09}{\GeV}$,
the combined ATLAS and CMS measurement with LHC Run-1 data~\cite{HIGG-2014-14}, which is consistent with the subsequent Run-2 measurements~\cite{HIGG-2022-20,
CMS-HIG-21-019, CMS-HIG-19-004}.

\subsection{Background modelling}

The $H\to \gamma\gamma$ process contributes as a resonant background
to the electron categories via photons misidentified as electrons.
It is modelled using the same DSCB functional form as the $H\to\gamma^{*}\gamma$ signal.
The limited event yields are insufficient to independently determine the shape parameters;
these are therefore fixed to those of the corresponding signal model.
The expected yields are determined separately in each category using the simulation
normalised to the predicted SM $H\to\gamma\gamma$ cross-section times branching ratio.
The background is predicted to be, on average, 2.4\% and 16\% of the expected
$H\to\gamma^{*}\gamma \to \ell\ell\gamma$ signal
in the resolved and merged dielectron channels, respectively.

Several processes contribute to the non-resonant background:
prompt $\ell\ell\gamma$ production and events with fake leptons or photons from non-prompt sources and misidentified jets.
The normalisation and shape of this background are determined from data in the final fit.
Templates built from simulation of the prompt $\ell\ell\gamma$ background together with non-prompt contributions estimated
with a data-driven technique are used to select the analytical functional form and evaluate the spurious signal uncertainty.

\subsubsection{Non-resonant background composition}
\label{sec:bkg_composition}

The fraction of events with fake photons is estimated in a
control region with the same selections as the signal region
except for the photon isolation requirement,
allowing the use of the photon isolation distribution to
determine prompt and fake contributions.
A background template
describing events with fake photons is created using data for each of the nine
categories by inverting the photon \textit{Tight} identification
requirement. Prompt photon contamination is removed
from the template using simulated events.
The resulting fake photon fraction in the signal regions is estimated
to be approximately 5\% across all categories.

The fraction of events containing fake leptons is
estimated in an analogous way. A template fit to the
subleading-lepton isolation variable is performed to extract the
fractions of fake leptons for each category. The estimated fractions
of events with fake leptons are up to 20\% for the electron
categories and 3\% for the muon categories.
The fake fractions estimated in the low $p_\mathrm{Tt}$ categories are also
applied in the VBF categories, as these do not have enough events to perform
the procedure.

\subsubsection{Non-resonant background template}
The prompt non-resonant background templates are derived from the
simulated non-resonant
$\ell\ell\gamma$ samples described in Section~\ref{sec:samples} for
each category separately.
The $m_{\ell\ell\gamma}$ distributions of the non-resonant background with fake photons or leptons
is derived from control regions (one per category)
created by inverting the object identification requirements.
Prompt photon and lepton contamination is removed using simulated events.
The resulting fake photon and fake lepton $m_{\ell\ell\gamma}$ distributions in data
are found to be similar to those from the simulated prompt non-resonant events.

\begin{figure}[!htp]
\centering
\includegraphics[width=0.5\linewidth]{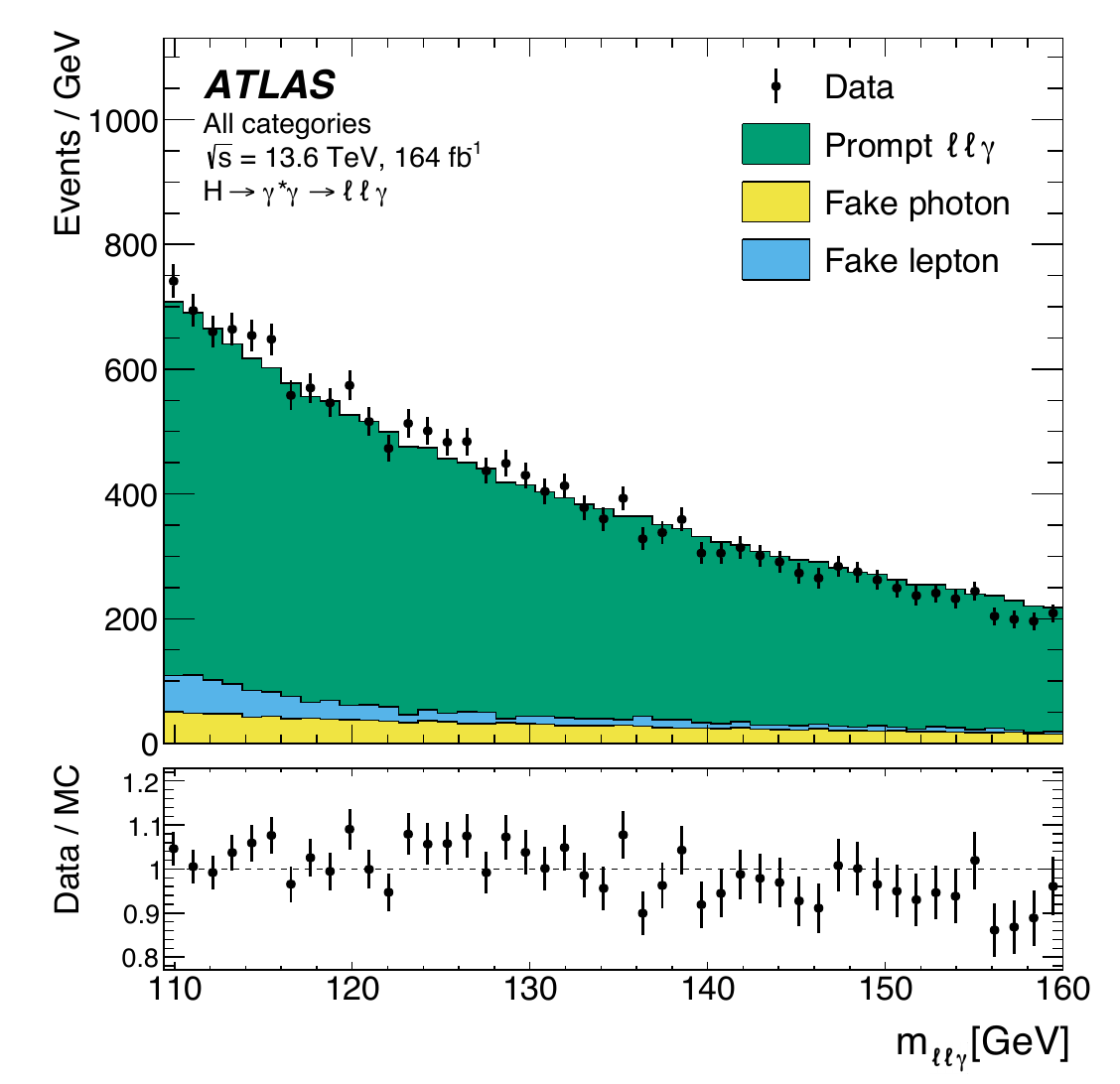}
\caption{The $m_{\ell\ell\gamma}$ distribution for events satisfying
the $H\to\gamma^{*}\gamma \to \ell\ell\gamma$
selection in data for all categories.
The data are shown as the black points with statistical uncertainties.
The contribution from the non-resonant $\ell\ell\gamma$ process (green),
the extra contributions from the fake processes (blue and yellow)
are shown for illustration.
The total background estimate is normalised to the number of data
events in the sideband.
The bottom panel shows the ratio of the data with respect to the
sum of background components.
}
\label{fig:bkg_compositions}
\end{figure}

The contributions from non-resonant $\ell\ell\gamma$, fake photon,
and fake lepton components are shown in
Figure~\ref{fig:bkg_compositions} combined across
all analysis categories.
Since all three non-resonant components (prompt $\ell\ell\gamma$, fake photon
and fake lepton) have similar $m_{\ell\ell\gamma}$ distributions,
the simulated prompt $\ell\ell\gamma$ sample is reweighted separately
in each category to match the total
non-resonant $m_{\ell\ell\gamma}$ yield and shape.
A first-order polynomial
($f(m_{\ell\ell\gamma}) = p_{0} +
p_{1}m_{\ell\ell\gamma}$) is used as a reweighting function for all
categories, except for the low $p_\mathrm{Tt}$ resolved $ee$ category, where
a linear-plus-exponential
($f(m_{\ell\ell\gamma}) = p_{0} + p_{1}m_{\ell\ell\gamma} +
p_{2}e^{-p_{3}m_{\ell\ell\gamma}}$) function is chosen based on a
$\chi^{2}/$ndf criterion.

\subsubsection{Non-resonant background function}
The reweighted $\ell\ell\gamma$ sample, accounting for prompt,
fake photon and lepton contributions, is used to choose a
function to describe the non-resonant background in each category.

The  function to represent the background without artificial signal
features is determined in a spurious signal study~\cite{HIGG-2018-42}.
In each category, an analytic function is chosen by fitting them to the background-only
template in the mass range of
$\qty{110}{\GeV} < m_{\ell\ell\gamma} < \qty{160}{\GeV}$.
A function is chosen if it satisfies two criteria:
\begin{itemize}
\item The spurious signal, $|SS|$, is defined as the absolute value of the
maximum fitted signal yield~\cite{ATL-PHYS-PUB-2020-028}.
The procedure is repeated by varying
the Higgs mass between \qty{120}{\GeV} and \qty{130}{\GeV} in
steps of \qty{1}{\GeV}.
The spurious signal must be less than 20\% of the statistical uncertainty
on the fitted signal yield, or 10\% of the expected signal yield.
\item The $\chi^{2}$ probability must be above 1\%.
\end{itemize}
When more than one function passes these criteria, the
one with the fewest free parameters is chosen.
If more than one has the same number of parameters, the one with the
smallest |\textit{SS}| is chosen.
Following this procedure, an exponential of a first-order polynomial function is
chosen for most categories.
A power-law function is used for the $\mu\mu$ low $p_\mathrm{Tt}$ and
$ee$ resolved high $p_\mathrm{Tt}$ categories,
a first-order dijet-like function~\cite{EXOT-2010-01} is used for
the $ee$ resolved low $p_\mathrm{Tt}$ category,
and a second-order Bernstein polynomial function is selected for the $ee$ merged low $p_\mathrm{Tt}$ category.
The overall normalisation and function parameters are determined in
a fit to the data.

\subsection{Total fit model}

An unbinned extended likelihood function is constructed as the
product of the parameterized signal-plus-background probability
density functions across all categories.
Systematic uncertainties are incorporated as nuisance parameters with
Gaussian or log-normal constraints.
These are treated as correlated between the categories, except for
the spurious-signal uncertainties.
The latter are included independently in each category as a
signal-like component, with a yield constrained by a Gaussian
function centred at zero and with a width corresponding to the
estimated spurious signal.

The parameter of interest is the signal strength
$\mu$~\cite{Cowan:2010js}, defined as the ratio of the measured
signal yield to the SM prediction.
The best-fit value for $\mu$ is obtained by maximising the
corresponding profile likelihood ratio~\cite{Cowan:2010js}.
Any possible excess over the background-only hypothesis is quantified
by a \textit{p}-value using the profile likelihood ratio,
evaluated under the assumption of a vanishing $H\to
\ell\ell\gamma$ branching ratio.
A summary of the expected signal and background yields in the
smallest window containing 90\% of signal events is shown in
Table~\ref{tab:sandb_yields}.
\begin{table}[!htp]
\centering
\caption{
Expected signal ($S_{90}$) and background ($B_{90}$) events in the smallest
window containing 90\% of the signal events in each category.
$B_{90}$ is obtained by integrating the fitted background functions
over the 90\% signal window.
$H\to\gamma\gamma_{90}$ refers to the expected number of selected
$H\to\gamma\gamma$ events in the same window;
the symbol ``--'' indicates a negligible contribution in the corresponding categories.
The signal fraction $f_{90}$ and expected significance $Z_{90}$
are defined as $f_{90} = S_{90} / (S_{90}+B_{90})$ and
$Z_{90} = \sqrt{2((S_{90}+B_{90}) \ln (1+S_{90}/B_{90}) - S_{90})}$. }
\resizebox{\linewidth}{!}{
\begin{tabular}{lcrrr|rr}
\toprule
Category                             &
$[\mu_{\text{CB}}-\sigma_{90},\mu_{\text{CB}}+\sigma_{90}]$&
$S_{90}$  & $B_{90}$  & $H\to\gamma\gamma_{90}$  &
$f_{90}$[\%]   & $Z_{90}$  \\
\midrule
$\mu\mu$ low-$p_\mathrm{Tt}$       & [121.3, 128.7] & 57.0 & 1778 & --   & 3.1   & 1.4      \\
$ee$ resolved low-$p_\mathrm{Tt}$  & [120.9, 129.4] & 22.2 & 1034 & 0.79 & 2.1   & 0.7      \\
$ee$ merged low-$p_\mathrm{Tt}$    & [121.0, 128.8] & 36.7 & 1459 & 6.80 & 2.4   & 1.0      \\
$\mu\mu$ VBF                       & [121.0, 129.0] & 1.58 & 6.8  & --   & 19    & 0.6      \\
$ee$ resolved VBF                  & [121.6, 128.7] & 0.53 & 3.6  & 0.01 & 13    & 0.3      \\
$ee$ merged VBF                    & [121.7, 128.2] & 1.19 & 3.3  & 0.21 & 25    & 0.6      \\
$\mu\mu$ high-$p_\mathrm{Tt}$      & [121.1, 129.0] & 5.7  & 41.4 & --   & 12    & 0.9      \\
$ee$ resolved high-$p_\mathrm{Tt}$ & [121.8, 128.2] & 1.65 & 16.7 & 0.02 & 9.0   & 0.4      \\
$ee$ merged high-$p_\mathrm{Tt}$   & [122.1, 128.1] & 4.6  & 28.8 & 0.54 & 13    & 0.8      \\
\bottomrule
\end{tabular}
}
\label{tab:sandb_yields}
\end{table}


%

%

\section{Systematic uncertainties}
\label{sec:systematic}

Systematic uncertainties in the signal modelling arise from
both theory and experimental sources,
and are incorporated into the signal-extraction fit via nuisance parameters
that affect the signal yields and mass distributions in each category.
Signal shape uncertainties vary by category and are due to the calibration
of the electron and photon energy and muon momentum.
The uncertainty in the electron and photon energy resolution and in
the muon momentum resolution leads to an average uncertainty of 5.8\% in the mass
resolution ($\sigma_{\text{CB}}$).
The uncertainty in the electron and photon energy scale (and in the
muon momentum scale) leads to 0.3\% (0.4\%) uncertainty
in the peak position ($\mu_{\text{CB}}$).

Signal normalisation uncertainties also vary by category.
As discussed in Section~\ref{sec:samples}, the total trigger efficiency determined by simulation
is 92\% for electron and muon events passing all offline selections.
To account for possible differences in trigger efficiencies between data and simulation,
a 5\% uncertainty is assigned to all events.
An additional 15\% uncertainty is assigned to events in the $\mu\mu$
categories, as the Level-1 single-muon trigger in the barrel region
is known to be less efficient in data than simulation by 10--20\%~\cite{TRIG-2018-01, TRIG-2019-02, TRIG-2022-01},
while multi-muon trigger scale factors are generally better modelled.
Therefore, a conservative 15\% uncertainty is assigned to cover these deviations in the logical OR of the single- and multi-object triggers.
Uncertainties in the signal yield arising from
reconstruction, identification, and isolation efficiencies are around 1\%
for photons, electrons, and muons.
An additional 20\% uncertainty is assigned in the merged dielectron categories to account for the uncertainty in the BDT efficiency.
The jet-related uncertainties, including those from jet
reconstruction and in situ calibration, are below 3\%.
The uncertainty in the combined 2022--2024 integrated luminosity is
$\pm 1.9\%$~\cite{ATL-DAPR-PUB-2025-001}, obtained using the LUCID-2 detector~\cite{LUCID2}
for the primary luminosity measurements.

Theory uncertainties affecting the expected signal yield arise from
multiple sources.
Uncertainties in the production cross-section and kinematic
distributions, primarily due to missing higher-order QCD corrections,
contribute up to 16\% depending on the analysis category.
An additional QCD scale uncertainty is assigned to account for the migration of ggF events
across the VBF categories~\cite{HIGG-2018-43}.
The parton shower modelling uncertainty
ranges from 1\% to 23\%, with the largest effects observed in the
high $p_\mathrm{Tt}$ categories.
The uncertainty in the $H\to\gamma^{*}\gamma
\to \ell\ell\gamma$ branching ratio is taken as
6.9\%~\cite{HIGP-2024-19}.

These theory uncertainties are also considered for the
$H\to\gamma\gamma$ background normalisation, alongside two
additional uncertainties: a 9.4\% uncertainty associated with the
precision of the measured cross-section~\cite{HIGG-2018-29} and a
50\% uncertainty to account for potential mismodelling of
electron-to-photon fake-rates~\cite{PERF-2017-02}.

The spurious signal uncertainty is applied as an uncertainty on the
signal yield and accounts for uncertainty in the description of the
non-resonant background.
Table~\ref{tab:syst_breakdown} shows the breakdown of the impact
of individual systematic uncertainty sources on both the expected
and the observed signal strength.

\begin{table}[!htp]
\centering
\caption{Breakdown of the impact of individual systematic
uncertainty sources on the expected and observed signal strength.
The ``Others'' category includes uncertainties from pile-up modelling, the $H\to\gamma\gamma$ cross-section and fake-rate estimates, and the $b\bar{b}H$ cross-section.}
\begin{tabular}{l c c}
\toprule
Uncertainty source                            & Expected ($\Delta \mu$) & Observed ($\Delta \mu$) \\
\midrule
Spurious signal (background modelling)        & 0.04                    & 0.04  \\
QCD scale, PDF+\alphas, parton shower         & 0.07                    & 0.05  \\
Branching ratio                               & 0.07                    & 0.05  \\
Luminosity                                    & 0.02                    & 0.01  \\
Photon efficiencies                           & 0.06                    & 0.04  \\
Jet                                           & 0.01                    & 0.01  \\
Electron efficiency                           & 0.02                    & 0.01  \\
Merged electron ID                            & 0.07                    & 0.06  \\
Electron/photon energy scale and resolution   & 0.03                    & 0.03  \\
Muon                                          & 0.08                    & 0.05  \\
Others                                        & 0.03                    & 0.03  \\
\midrule
Systematic uncertainty                        & 0.17                    & 0.13  \\
Statistical uncertainty                       & 0.40                    & 0.39  \\
\midrule
Total                                         & 0.44                    & 0.42  \\
\bottomrule
\end{tabular}
\label{tab:syst_breakdown}
\end{table}


%

%

\section{Results}
\label{sec:results}

The results are extracted through an unbinned, simultaneous
maximum-likelihood fit~\cite{Cowan:2010js} to the
$m_{\ell\ell\gamma}$ distributions in all categories.
The likelihood function is built by incorporating both the signal
strength of $H\rightarrow \gamma^{*}\gamma$ and the nuisance
parameters described in Section~\ref{sec:systematic},
which characterise the effects of systematic uncertainties on the
signal normalisation and shape, as well as background normalisation
and shape parameters.
The best-fit value of the signal strength parameter is $\mu =
0.64^{+0.44}_{-0.39} = 0.64^{+0.40}_{-0.39} (\mathrm{stat.})
^{+0.18}_{-0.08} (\mathrm{syst.})$ for $m_{H} = \qty{125.09}{\GeV}$.
Under the SM signal hypothesis, the expected signal strength is $\mu
= 1.00^{+0.47}_{-0.41} = 1.00^{+0.41}_{-0.40} (\mathrm{stat.})
^{+0.23}_{-0.11} (\mathrm{syst.})$.

\begin{figure}[!htp]
\centering
\includegraphics[width=0.5\linewidth]{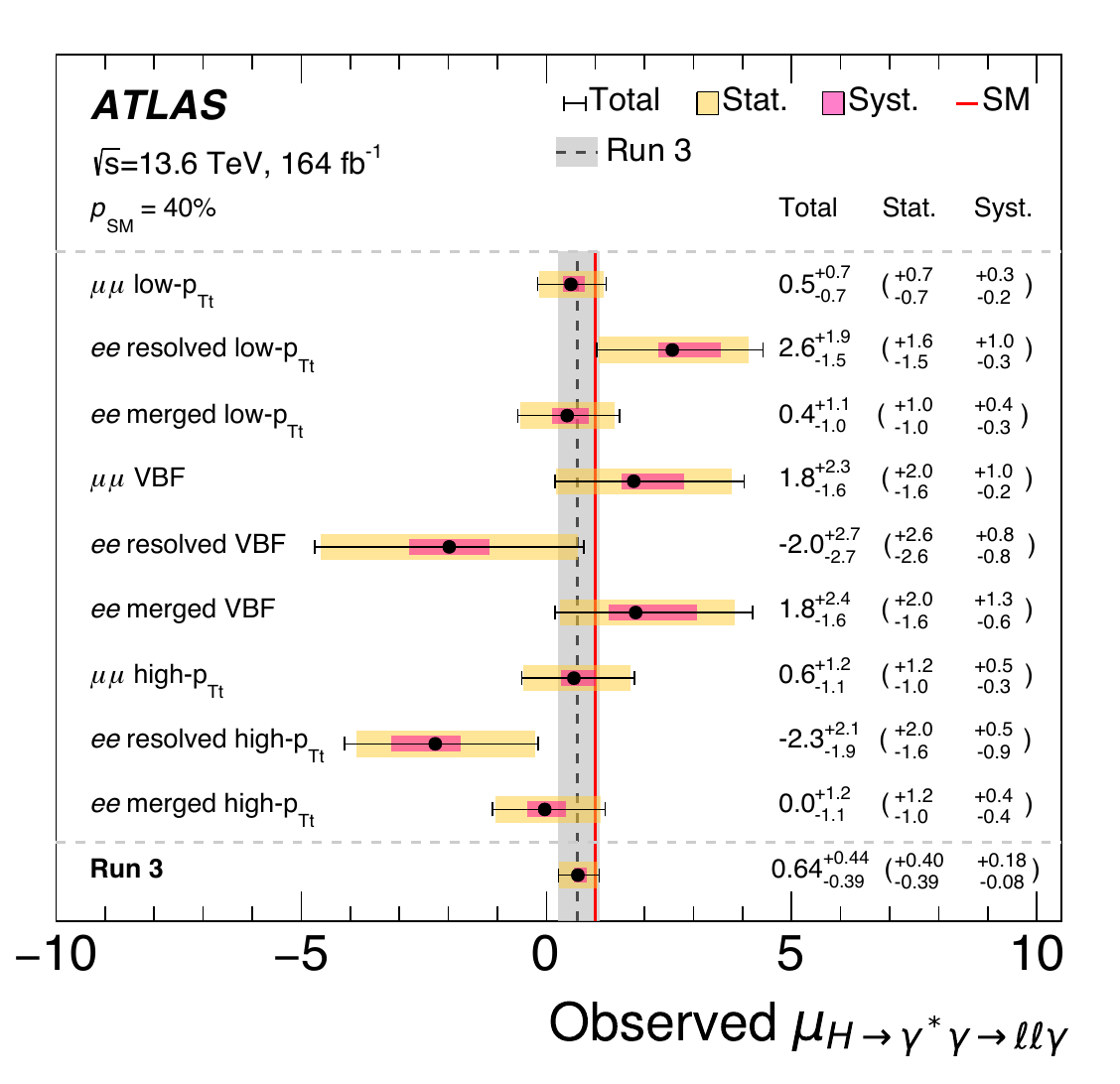}
\caption{Observed best-fit values of the signal strengths and
corresponding uncertainties in each analysis category, in a fit
where the signal strength in each category is allowed to float
independently,
compared with the result of the Run-3 global fit (last line,
black vertical line)
including its total uncertainty (grey band) and with the SM (red
vertical line).
The error bars correspond to the total uncertainties and the
coloured bands indicate the statistical
and systematic uncertainties.}
\label{fig:category_mu_observed}
\end{figure}

\begin{figure}[!htp]
\centering
\subfloat[]{\includegraphics[width=0.35\textwidth]{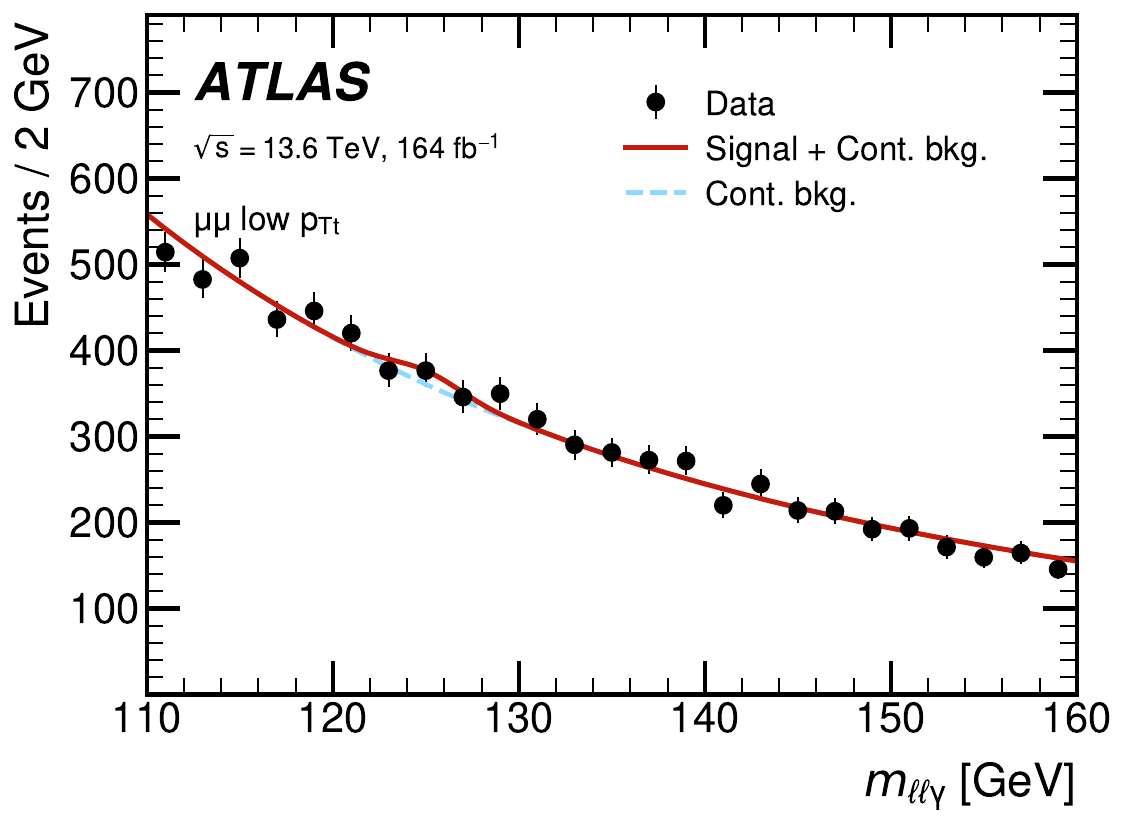}}
\subfloat[]{\includegraphics[width=0.35\textwidth]{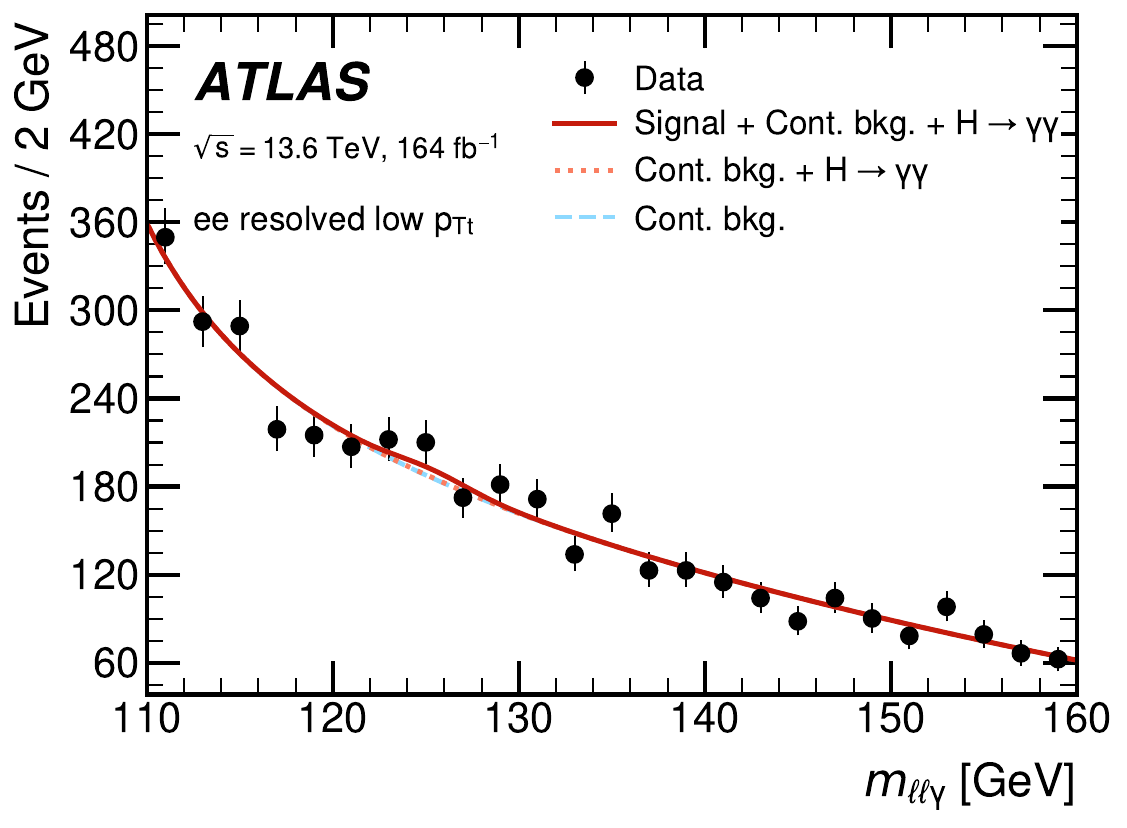}}
\subfloat[]{\includegraphics[width=0.35\textwidth]{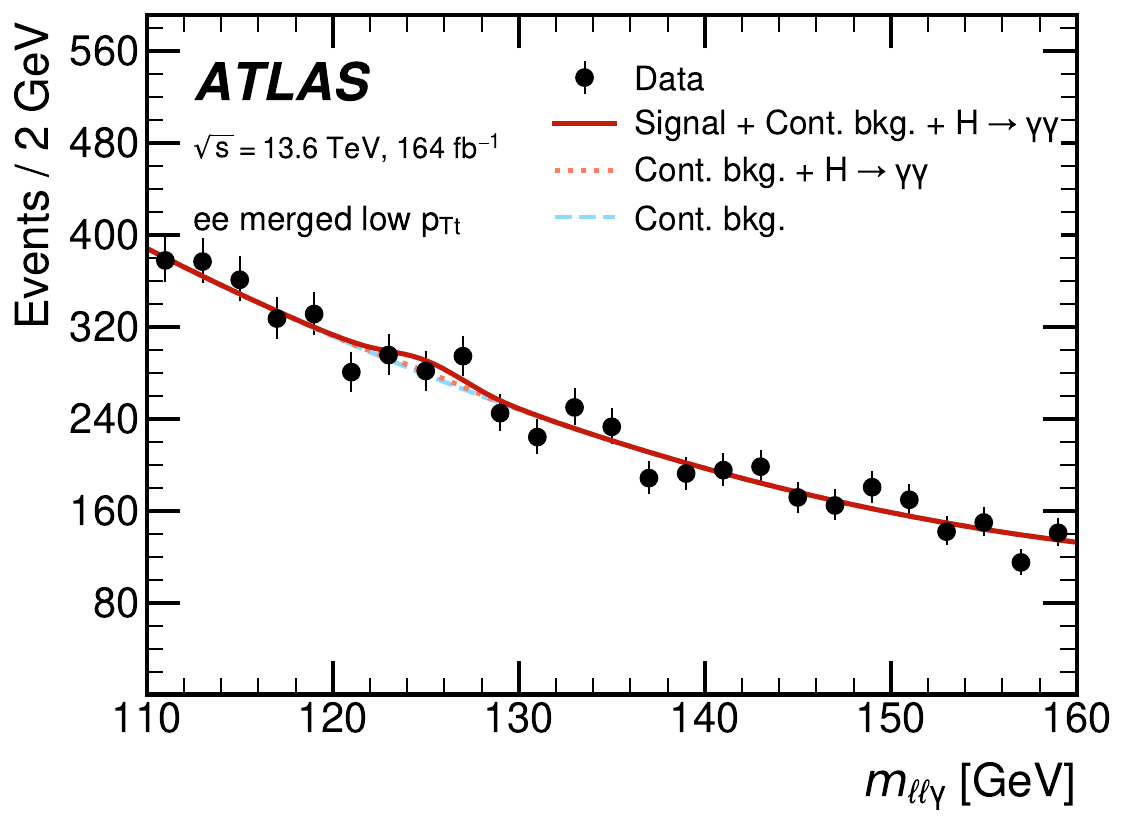}}
\\
\subfloat[]{\includegraphics[width=0.35\textwidth]{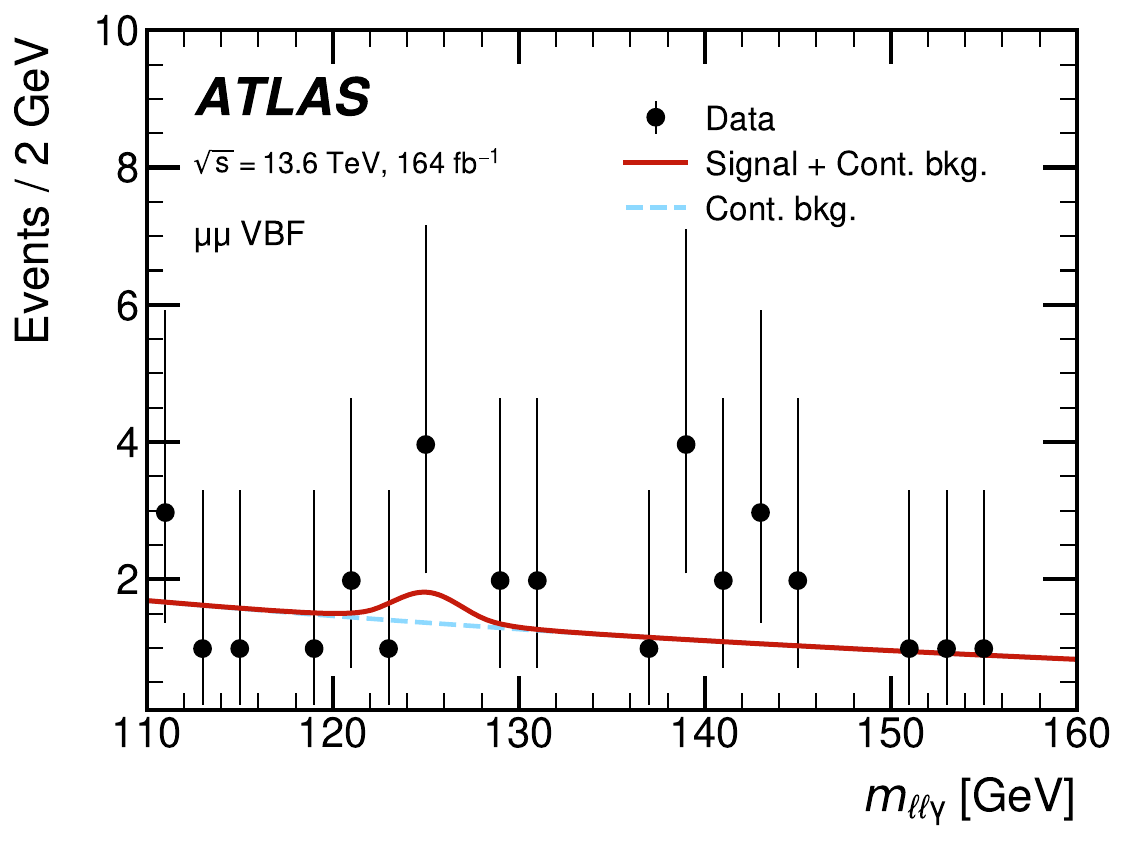}}
\subfloat[]{\includegraphics[width=0.35\textwidth]{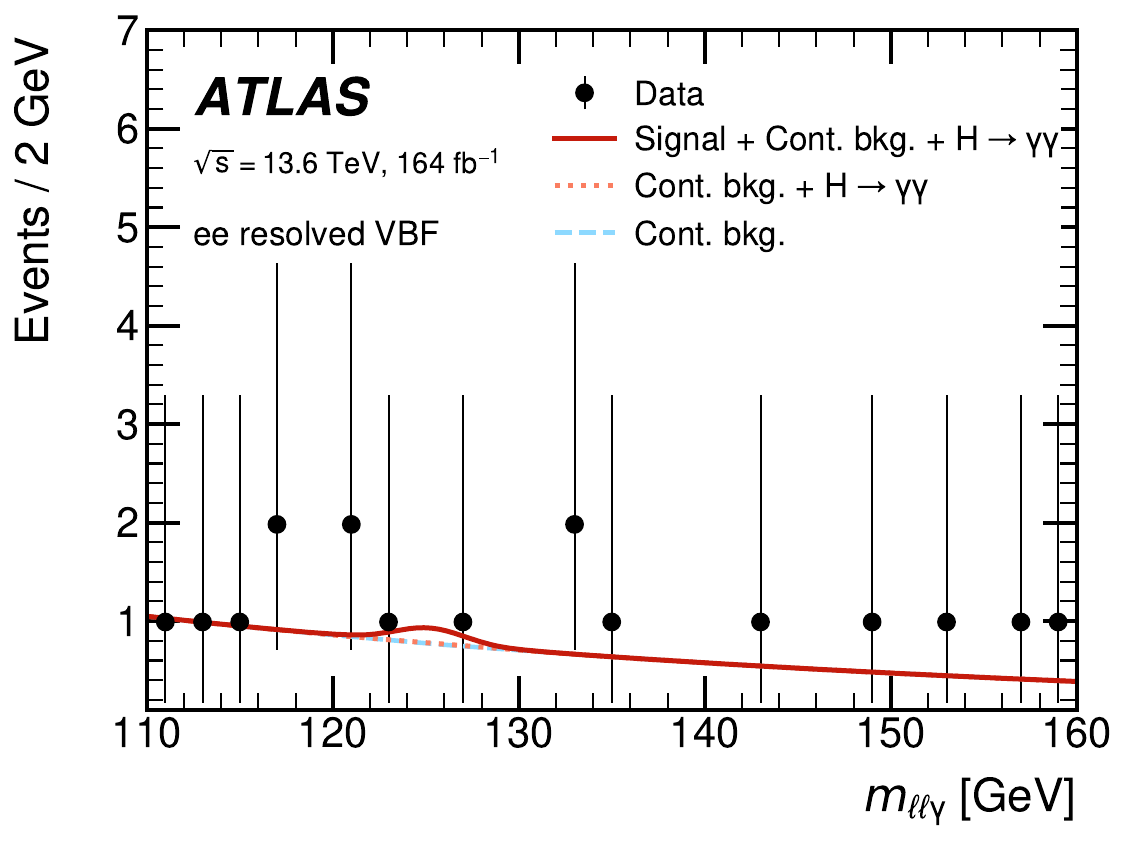}}
\subfloat[]{\includegraphics[width=0.35\textwidth]{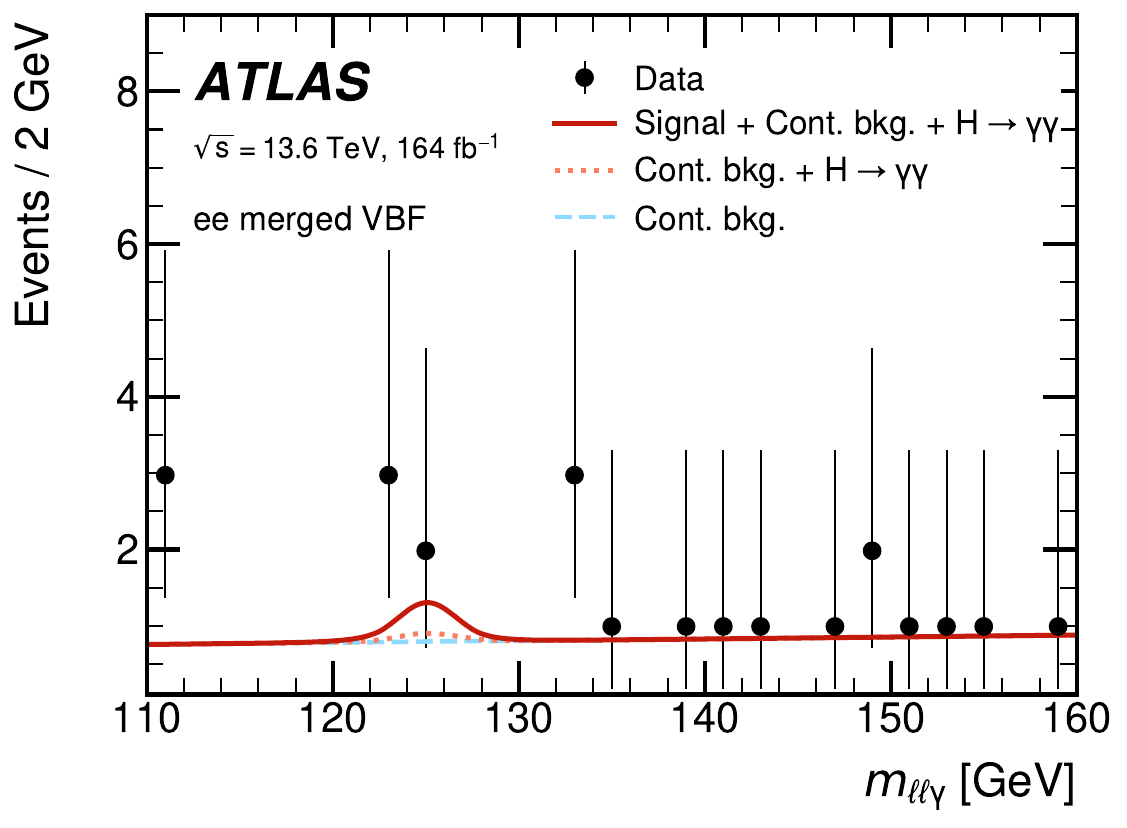}}
\\
\subfloat[]{\includegraphics[width=0.35\textwidth]{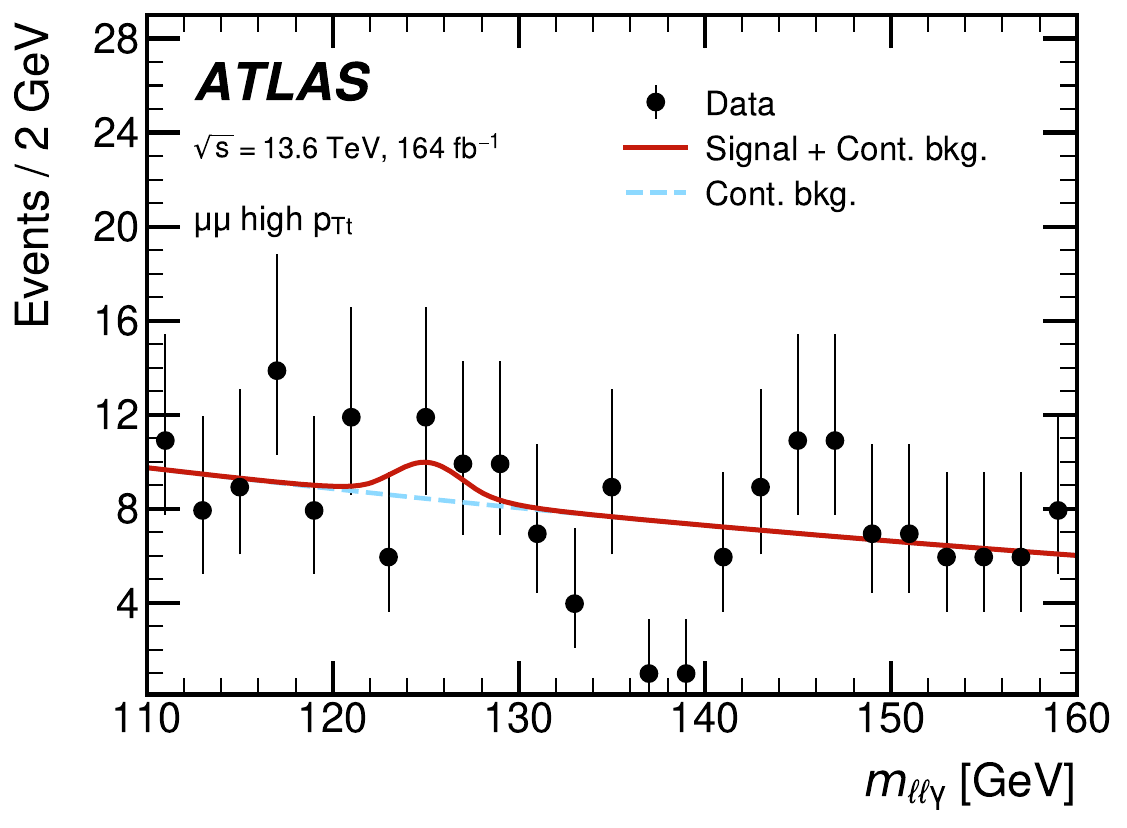}}
\subfloat[]{\includegraphics[width=0.35\textwidth]{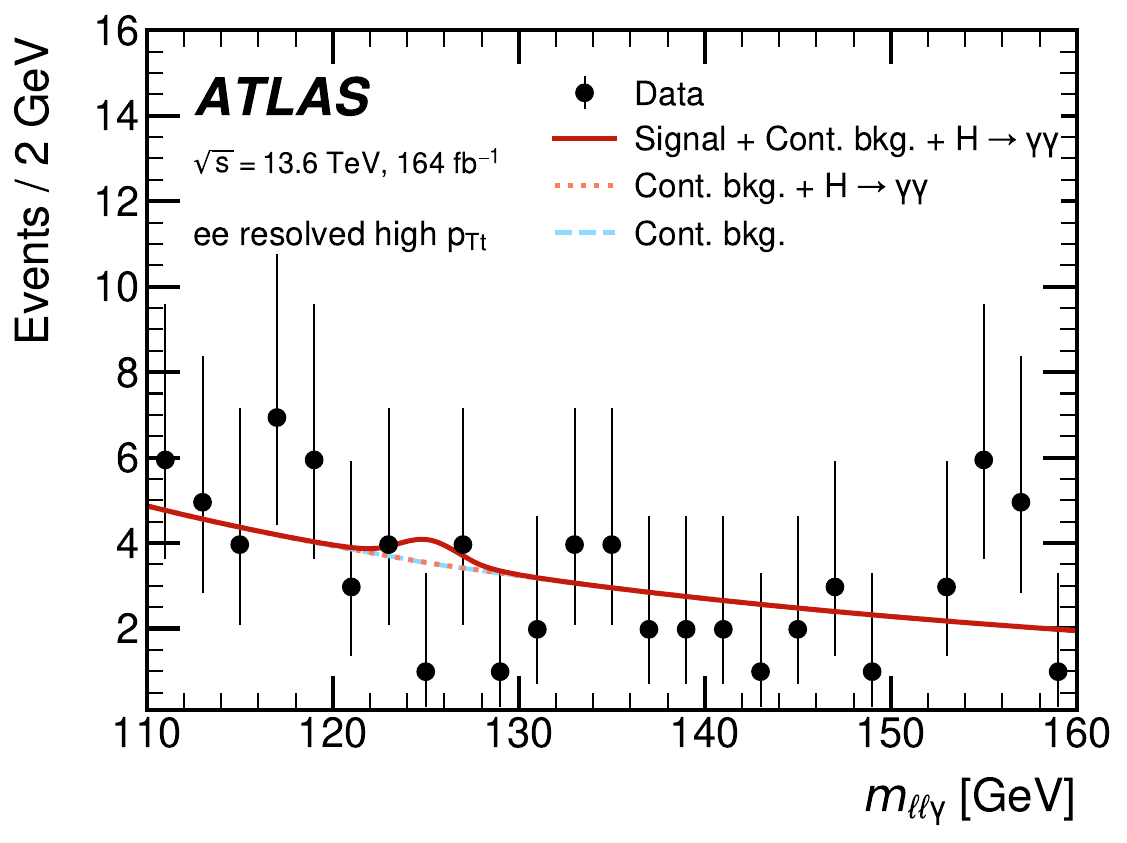}}
\subfloat[]{\includegraphics[width=0.35\textwidth]{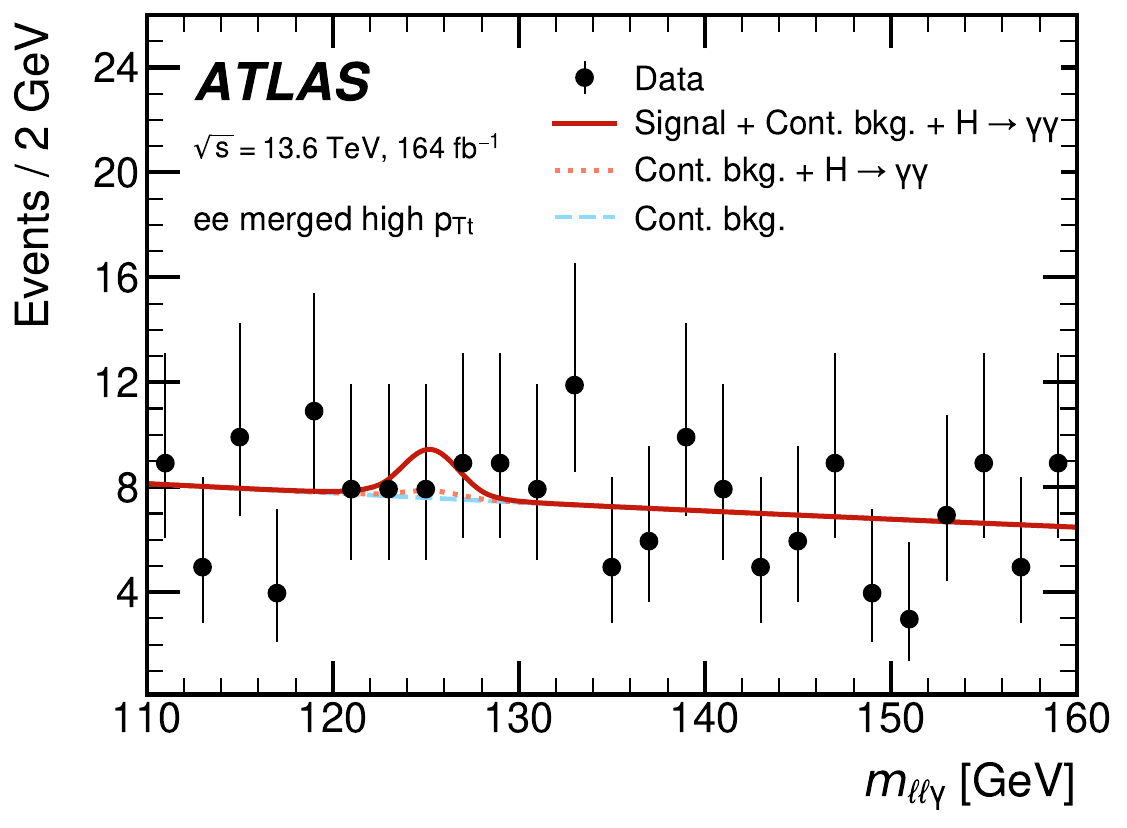}}
\\
\caption{Invariant mass distributions $m_{\ell\ell\gamma}$ of the
selected events and the results of the global fit, for the~(a,b,c)
low-$p_\mathrm{Tt}$ categories,
the~(d,e,f) VBF categories, and the~(g,h,i) high-$p_\mathrm{Tt}$ categories.
The $\mu\mu\gamma$ categories are shown in the left column, the
$ee\gamma$ resolved categories are in the middle, and the
$ee\gamma$ merged categories are in the right column.
The data points are shown as the black points with statistical
uncertainties.
The blue dashed curve represents the continuum background,
the dotted orange curve represents the continuum background plus the
resonant $H\rightarrow\gamma\gamma$ contribution where applicable,
and the red curve shows the signal-plus-background prediction.}
\label{fig:splusb_fit}
\end{figure}

\begin{figure}[!htp]
\centering
{\includegraphics[width=0.5\linewidth]{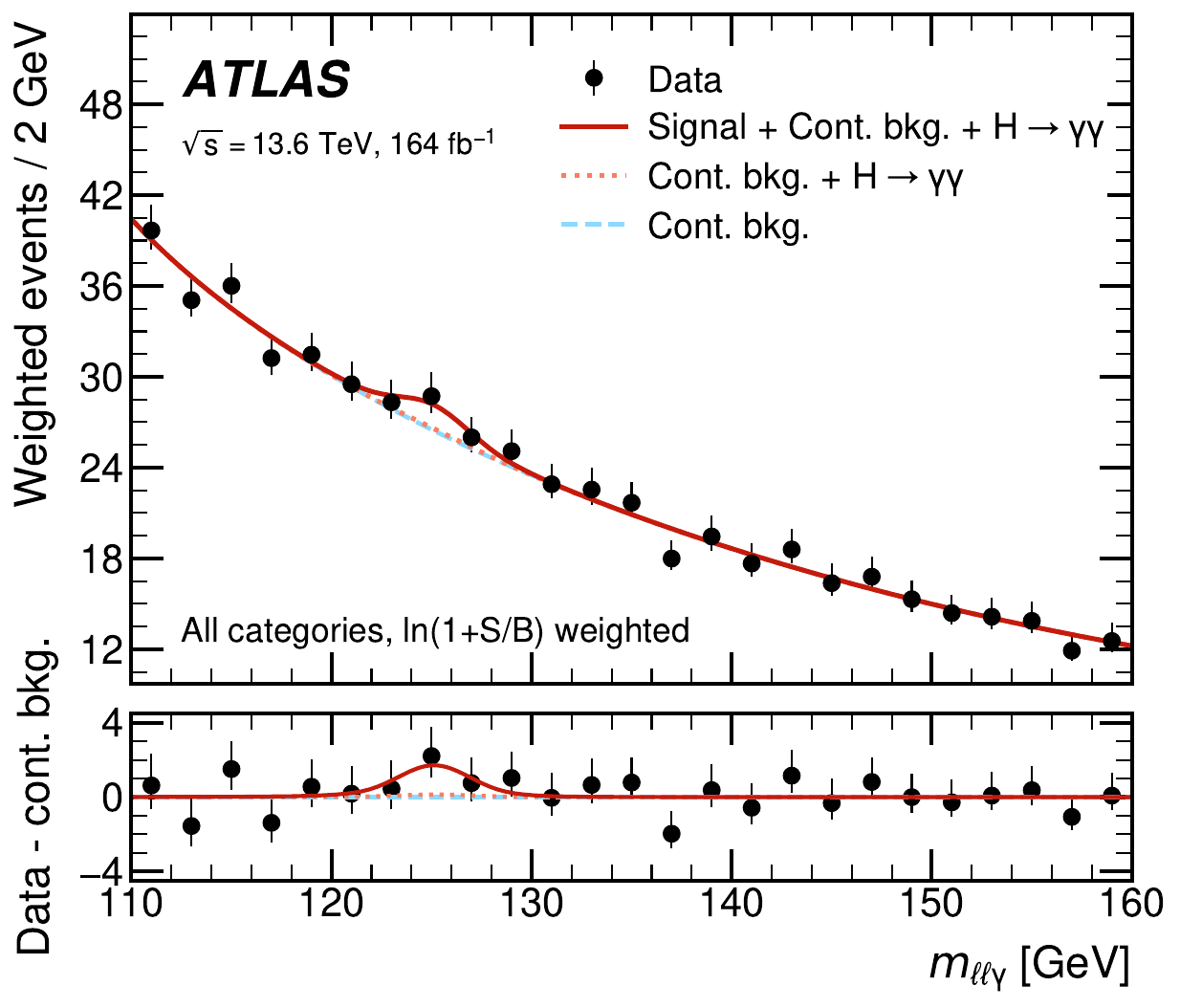}}
\caption{$m_{\ell\ell\gamma}$ distribution for the Run-3 data set,
with every event
reweighted by a category-dependent weight, $\ln(1+S_{90}/B_{90})$,
where $S_{90}$ is the number of signal events in the smallest
window containing 90\% of the expected signal
and $B_{90}$ is the expected number of background events in the
same window, estimated from fits to the data sidebands using the
background models.
The data points are shown as the black circles with statistical
uncertainties.
The parameterized signal and backgrounds are also added up with
the category-dependent weight.
The red curve shows the combined signal-plus-background model
when fitting all analysis categories simultaneously,
the dashed blue line shows the model of the non-resonant
background component and the dotted orange line denotes the sum of
the non-resonant background and the resonant
$H\rightarrow\gamma\gamma$ background.
The bottom panel shows the residuals of the data relative to the
non-resonant background component of the signal-plus-background fit.}
\label{fig:splusb_fit_weighted}
\end{figure}

Figure~\ref{fig:category_mu_observed} shows the observed
result of the fit when the signal strength in each category is
allowed to float independently. The combined result is also included.
The resulting measurement is compatible with the SM prediction, with a
$p$-value of 0.41. The $m_{\ell\ell\gamma}$ distribution in each
category, with the result of the global fit of the parametric
signal-plus-background models to the data, is shown in
Figure~\ref{fig:splusb_fit}.
Figure~\ref{fig:splusb_fit_weighted} shows the
$m_{\ell\ell\gamma}$ distribution with every data event reweighted
by a category-dependent weight $\ln(1+S_{90}/B_{90})$,
where $S_{90}$ and $B_{90}$ are the number of signal and background events in
the smallest window containing
90\% of the expected signal as listed in Table~\ref{tab:sandb_yields}.

The observed (expected) significance over the background-only
hypothesis for a Higgs boson with a mass of \qty{125.09}{\GeV} is
1.7 $\sigma$ (2.5 $\sigma$).
The Higgs boson production cross-section times the
$H\rightarrow\ell\ell\gamma$ branching ratio is determined to be $4.1^{+2.8}_{-2.5}$~fb.
\footnote{The treatment of signal theory uncertainties differs between
the signal strength and cross-section times branching ratio
($\sigma_H \times \mathcal{B}(H\to\ell\ell\gamma)$) measurements.
In the signal strength measurement, these uncertainties are evaluated
through the expected SM signal yields, whereas in the ($\sigma_H
\times \mathcal{B}$) measurement only the uncertainty on the signal
acceptance is considered.}

\subsection{Run-2 and Run-3 combination}
\begin{figure}[!htp]
\centering
\includegraphics[width=0.5\linewidth]{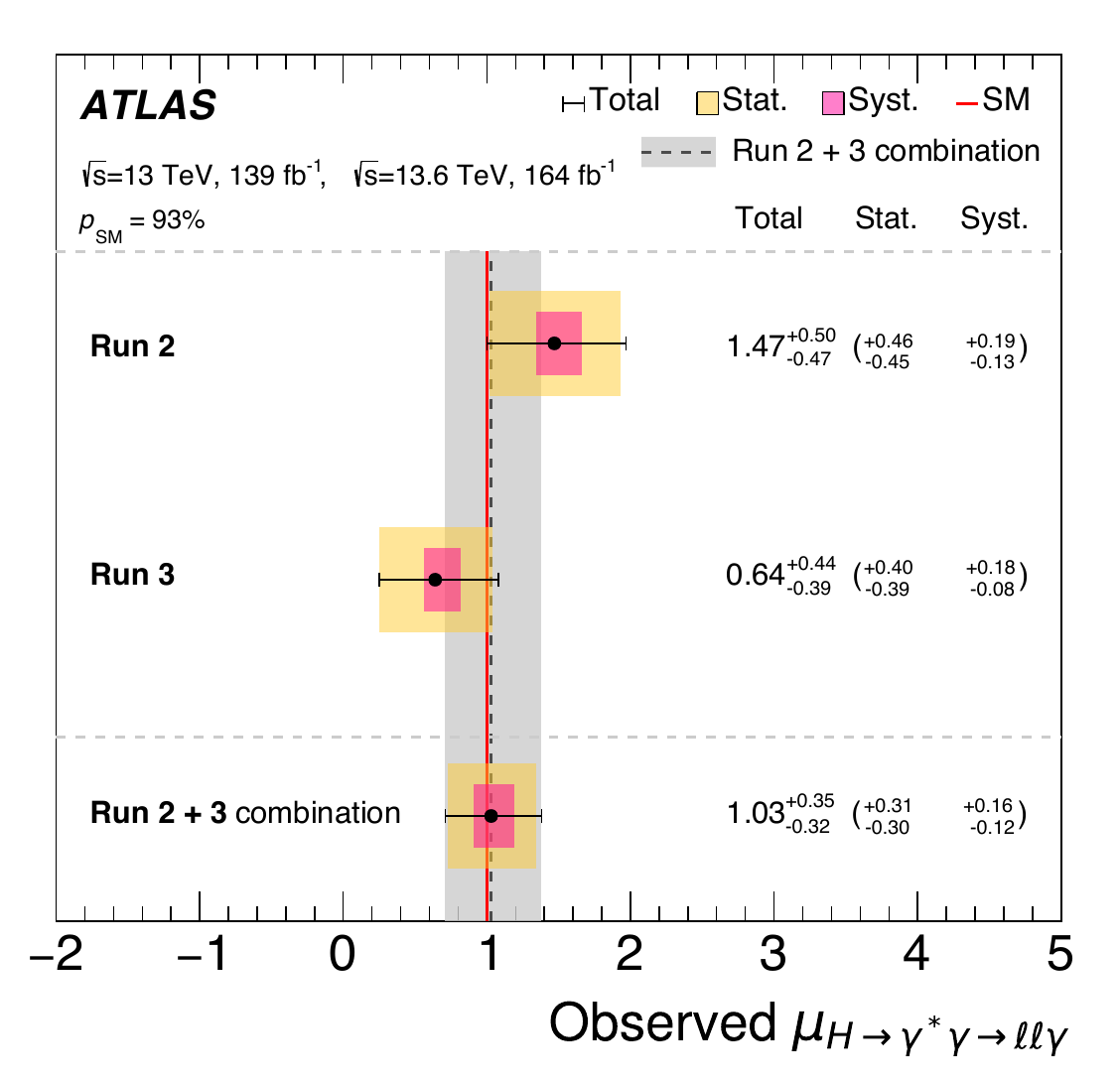}
\caption{The observed signal strengths from Run-2
(\qty{13}{\TeV}) and Run-3 (\qty{13.6}{\TeV}), and the combined
Run-2 and Run-3 results, with corresponding uncertainties.
The error bars correspond to the total uncertainties and the
coloured bands indicate the statistical and systematic uncertainties.}
\label{fig:fit_signal_strength}
\end{figure}

\begin{figure}[!htp]
\centering
\includegraphics[width=0.5\textwidth]{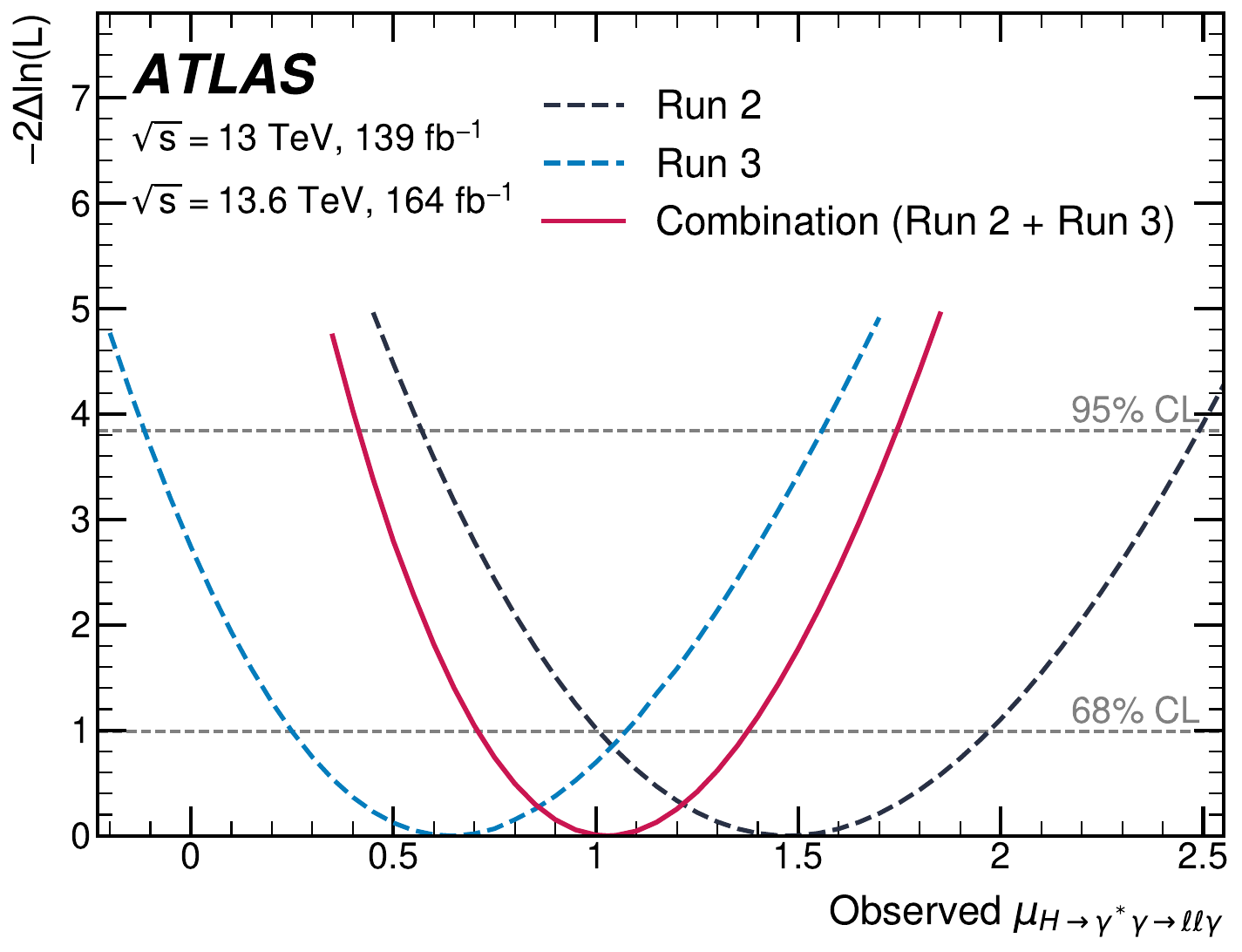}
\caption{Profiled likelihood scan of the signal strength for the
observed result.
The plot shows the results obtained by this analysis of the
Run-3 data (blue line),
by the previous analysis of the Run-2 data~\cite{HIGG-2018-43} (black line),
and their combination (red line).}
\label{fig:fit_likelihood_obs}
\end{figure}

The results presented in this paper are combined with the analysis of
the previous ATLAS search for $H\rightarrow\gamma^{*}\gamma$ with Run-2
data~\cite{HIGG-2018-43}.
The theory uncertainties are in general treated as fully correlated
across both data sets. However, the PDF uncertainties are treated as
uncorrelated
due to the different PDF sets
used in Run~2 and Run~3, and the QCD scale uncertainties are also
uncorrelated due to different implementations in the two analyses.
Considering the different data-taking conditions, most experimental
uncertainties are treated as uncorrelated between the two runs; in particular,
the uncertainties related to the trigger, luminosity, electron,
photon, and muon are treated as uncorrelated. Conversely,
uncertainties related to jets are correlated between the two results.
The joint likelihood is constructed as the product of the individual
likelihood functions from the Run-2 and Run-3 analyses.
Figure~\ref{fig:fit_signal_strength} shows the
observed signal strengths from both analyses and their combination.
Figure~\ref{fig:fit_likelihood_obs} shows the
likelihood scan of the combined signal strength.

The observed combined fit has a best-fit signal strength of $\mu =
1.03^{+0.31}_{-0.30} (\mathrm{stat.})^{+0.16}_{-0.12}
(\mathrm{syst.}) = 1.03^{+0.35}_{-0.32} (\mathrm{total})$,
with an expected $\mu = 1.00^{+0.31}_{-0.31}
(\mathrm{stat.})^{+0.16}_{-0.08} (\mathrm{syst.}) =
1.00^{+0.34}_{-0.32} (\mathrm{total})$.
The significance of the observed (expected) excess above the
background-only hypothesis is 3.4$\sigma$ (3.3$\sigma$).


%

%
%
%

%
\clearpage
\section{Conclusion}
\label{sec:conclusion}

The $H\rightarrow \gamma^{*}\gamma \rightarrow \ell\ell\gamma$ process is measured in the
$\mu\mu$, resolved electrons, and merged dielectron channels for $m_{\ell\ell} <$ \qty{30}{\GeV}.
The measurement uses proton--proton collision data recorded with the ATLAS detector at
$\sqrt{s} = \qty{13.6}{\TeV}$ during 2022--2024, corresponding to an integrated luminosity of
\qty{164}{\ifb}.
A newly developed BDT-based identification for merged dielectron objects improves
the overall expected significance by 5\% relative to the cut-based selection.

A simultaneous unbinned maximum-likelihood fit to the reconstructed
invariant mass across all event categories gives an observed
(expected) signal yield normalised
to the SM prediction, $\mu = 0.64^{+0.44}_{-0.39}$ ($\mu =
1.00^{+0.47}_{-0.41}$).
This corresponds to an observed (expected) signal significance of
1.7 (2.5) standard deviations.

This measurement is further combined with the Run-2 result,
improving the precision and sensitivity of the $H\rightarrow\gamma^{*}\gamma$ measurement.
In the combined fit, the observed (expected) signal yield normalised
to the SM prediction is $\mu =1.03^{+0.35}_{-0.32}$ ($\mu =
1.00^{+0.34}_{-0.32}$),
corresponding to an observed (expected) signal significance of 3.4
(3.3) standard deviations.

\section*{Acknowledgements}
%

%

%
%

%
%

We thank CERN for the very successful operation of the LHC and its injectors, as well as the support staff at
CERN and at our institutions worldwide without whom ATLAS could not be operated efficiently.

The crucial computing support from all WLCG partners is acknowledged gratefully, in particular from CERN, the ATLAS Tier-1 facilities at TRIUMF/SFU (Canada), NDGF (Denmark, Norway, Sweden), CC-IN2P3 (France), KIT/GridKA (Germany), INFN-CNAF (Italy), NL-T1 (Netherlands), PIC (Spain), RAL (UK) and BNL (USA), the Tier-2 facilities worldwide and large non-WLCG resource providers. Major contributors of computing resources are listed in Ref.~\cite{ATL-SOFT-PUB-2026-001}.

We gratefully acknowledge the support of ANPCyT, Argentina; YerPhI, Armenia; ARC, Australia; BMWFW and FWF, Austria; ANAS, Azerbaijan; CNPq and FAPESP, Brazil; NSERC, NRC and CFI, Canada; CERN; ANID, Chile; CAS, MOST, NSFC and FRFCU, China; Minciencias, Colombia; MEYS CR, Czech Republic; DNRF and DNSRC, Denmark; IN2P3-CNRS and CEA-DRF/IRFU, France; SRNSFG, Georgia; BMFTR, HGF and MPG, Germany; GSRI, Greece; RGC and Hong Kong SAR, China; ICHEP and Academy of Sciences and Humanities, Israel; INFN, Italy; MEXT and JSPS, Japan; CNRST, Morocco; NWO, Netherlands; RCN, Norway; MNiSW, Poland; FCT, Portugal; MNE/IFA, Romania; MSTDI, Serbia; MSSR, Slovakia; ARIS and MVZI, Slovenia; DSI/NRF, South Africa; MICIU/AEI, Spain; SRC and Wallenberg Foundation, Sweden; SERI, SNSF and Cantons of Bern and Geneva, Switzerland; NSTC, Taipei; TENMAK, T\"urkiye; STFC/UKRI, United Kingdom; DOE and NSF, United States of America.

Individual groups and members have received support from BCKDF, CANARIE, CRC and DRAC, Canada; CERN-CZ and PRIMUS, Czech Republic; COST, ERC, ERDF, Horizon 2020 and Marie Sk{\l}odowska-Curie Actions, European Union; Investissements d'Avenir Labex, Investissements d'Avenir Idex and ANR, France; DFG and AvH Foundation, Germany; Herakleitos, Thales and Aristeia programmes co-financed by EU-ESF and the Greek NSRF, Greece; BSF-NSF and MINERVA, Israel; NCN and NAWA, Poland; La Caixa Banking Foundation, CERCA and AGAUR programs from Generalitat de Catalunya and PROMETEO and GenT Programmes Generalitat Valenciana, Spain; G\"{o}ran Gustafssons Stiftelse, Sweden; The Royal Society and Leverhulme Trust, United Kingdom; Eric and Wendy Schmidt Fund for Strategic Innovation, United States of America.

In addition, individual members wish to acknowledge support from Chile: Agencia Nacional de Investigaci\'on y Desarrollo (ANID FONDECYT reg. 1230987, FONDECYT 1230812, FONDECYT 1240864, Fondecyt 3240661, Fondecyt Regular 1240721); China: Fundamental Research Funds for the Central Universities (010-63263105), Chinese Ministry of Science and Technology (MOST-2023YFA1605700, MOST-2023YFA1609300), National Natural Science Foundation of China (NSFC 12275265, NSFC-W2543005); Czech Republic: Czech Science Foundation (GACR - 24-11373S), Ministry of Education Youth and Sports (ERC-CZ-LL2327, FORTE CZ.02.01.01/00/22\_008/0004632); EU: H2020 European Research Council (ERC - 101002463); European Union: European Research Council (BARD No. 101116429, ERC - 101219398, ERC 101089007), European Regional Development Fund (HE COFUND GA No.101081355, ERDF), Marie Sklodowska-Curie Actions (GAP-101168829); France: Agence Nationale de la Recherche (ANR-22-EDIR-0002, ANR-24-CE31-0504-01); Germany: Deutsche Forschungsgemeinschaft (DFG - 469666862); China: Research Grants Council (GRF); Italy: Istituto Nazionale di Fisica Nucleare (LHC-MIUR - 28003/2025), Ministero dell'Università e della Ricerca (NextGenEU I53D23000820006 M4C2.1.1, SOE2024\_0000023); Japan: Japan Society for the Promotion of Science (JSPS KAKENHI  JP25H0063, JSPS KAKENHI JP22H04944, JSPS KAKENHI JP24K23939, JSPS KAKENHI JP24KK0251, JSPS KAKENHI JP25H00650, JSPS KAKENHI JP25H01291, JSPS KAKENHI JP25K01011, JSPS KAKENHI JP25K01023, JSPS KAKENHI JP25KK0047); Poland: Polish National Science Centre (NCN 2021/42/E/ST2/00350, NCN OPUS 2023/51/B/ST2/02507, NCN OPUS nr 2022/47/B/ST2/03059, NCN UMO-2019/34/E/ST2/00393, UMO-2022/47/O/ST2/00148, UMO-2023/49/B/ST2/04085, UMO-2023/51/B/ST2/00920, UMO-2024/53/N/ST2/00869); Spain: Agència de Gestió d'Ajuts Universitaris i de Recerca. (AGAUR - 2023 BP 00141), Ministry of Science and Innovation (RYC2019-028510-I, RYC2020-030254-I, RYC2021-031273-I, RYC2022-038164-I), Ministerio de Ciencia, Innovación y Universidades/Agencia Estatal de Investigaci\'on (EU NextGenerationEU (PRTR-C17.I1), PID2022-142604OB-C22); Sweden: Carl Trygger Foundation (Carl Trygger Foundation CTS 22:2312), Swedish Research Council (Swedish Research Council 2023-04654, VR 2021-03651, VR 2022-03845, VR 2022-04683, VR 2023-03403, VR 2024-05451, VR 2025-05940), Knut and Alice Wallenberg Foundation (KAW 2023.0366); United Kingdom: The Binks Trust, Royal Society (NIF-R1-231091); United States of America: U.S. Department of Energy (ECA DE-AC02-76SF00515), John Templeton Foundation (John Templeton Foundation 63206), Neubauer Family Foundation.

%
%


%
%
%
%
%
%

%
%
%

%
%
\printbibliography
\clearpage
 
\begin{flushleft}
\hypersetup{urlcolor=black}
{\Large The ATLAS Collaboration}

\bigskip

\AtlasOrcid[0000-0002-6665-4934]{G.~Aad}$^\textrm{\scriptsize 102}$,
\AtlasOrcid[0000-0001-7616-1554]{E.~Aakvaag}$^\textrm{\scriptsize 17}$,
\AtlasOrcid[0000-0002-5888-2734]{B.~Abbott}$^\textrm{\scriptsize 121}$,
\AtlasOrcid[0000-0002-0287-5869]{S.~Abdelhameed}$^\textrm{\scriptsize 83b}$,
\AtlasOrcid[0000-0002-1002-1652]{K.~Abeling}$^\textrm{\scriptsize 54}$,
\AtlasOrcid[0000-0001-5763-2760]{N.J.~Abicht}$^\textrm{\scriptsize 48}$,
\AtlasOrcid[0000-0002-8496-9294]{S.H.~Abidi}$^\textrm{\scriptsize 30}$,
\AtlasOrcid[0009-0003-6578-220X]{M.~Aboelela}$^\textrm{\scriptsize 44}$,
\AtlasOrcid[0000-0002-9987-2292]{A.~Aboulhorma}$^\textrm{\scriptsize 36e}$,
\AtlasOrcid[0000-0001-5329-6640]{H.~Abramowicz}$^\textrm{\scriptsize 154}$
% ...
\bigskip
\\

% ...
$^{17}$Department for Physics and Technology, University of Bergen, Bergen; Norway.\\
% ...
$^{30}$Physics Department, Brookhaven National Laboratory, Upton NY; United States of America.\\
% ...
$^{36}$$^{(a)}$Facult\'e des Sciences Ain Chock, Universit\'e Hassan II de Casablanca;$^{(b)}$Facult\'{e} des Sciences, Universit\'{e} Ibn-Tofail, K\'{e}nitra;$^{(c)}$Facult\'e des Sciences Semlalia, Universit\'e Cadi Ayyad, LPHEA-Marrakech;$^{(d)}$LPMR, Facult\'e des Sciences, Universit\'e Mohamed Premier, Oujda;$^{(e)}$Facult\'e des sciences, Universit\'e Mohammed V, Rabat;$^{(f)}$Institute of Applied Physics, Mohammed VI Polytechnic University, Ben Guerir; Morocco.\\
% ...
$^{44}$Physics Department, Southern Methodist University, Dallas TX; United States of America.\\
% ...
$^{48}$Fakult\"{a}t Physik, Technische Universit{\"a}t Dortmund, Dortmund; Germany.\\
% ...
$^{54}$II. Physikalisches Institut, Georg-August-Universit\"{a}t G\"ottingen, G\"ottingen; Germany.\\
% ...
$^{83}$$^{(a)}$Khalifa University of Science and Technology, Abu Dhabi;$^{(b)}$New York University Abu Dhabi, Abu Dhabi;$^{(c)}$United Arab Emirates University, Al Ain;$^{(d)}$University of Sharjah, Sharjah; United Arab Emirates.\\
% ...
$^{102}$CPPM, Aix-Marseille Universit\'e, CNRS/IN2P3, Marseille; France.\\
% ...
$^{121}$Homer L. Dodge Department of Physics and Astronomy, University of Oklahoma, Norman OK; United States of America.\\
% ...
$^{154}$Raymond and Beverly Sackler School of Physics and Astronomy, Tel Aviv University, Tel Aviv; Israel.\\
% ...
$^{a}$ Also at Affiliated with an institute formerly covered by a cooperation agreement with CERN.\\
$^{b}$ Also at An-Najah National University, Nablus; Palestine.\\
$^{c}$ Also at Borough of Manhattan Community College, City University of New York, New York NY; United States of America.\\
$^{d}$ Also at Center for Interdisciplinary Research and Innovation (CIRI-AUTH), Thessaloniki; Greece.\\
$^{e}$ Also at Centre of Physics of the Universities of Minho and Porto (CF-UM-UP); Portugal.\\
$^{f}$ Also at CERN, Geneva; Switzerland.\\
% ...

\end{flushleft}


%

%
%

%

%
%
%
%

\end{document}